\documentclass[twocolumn]{emulateapj09}

\usepackage{graphicx}
\usepackage{epsfig}

\newcommand{\sgra}{Sgr~A$^*$}

\shorttitle{Null Hypothesis GR Test with \sgra}
\shortauthors{Psaltis et al.}

\begin{document}


\title{A General Relativistic Null Hypothesis Test with Event Horizon
  Telescope Observations of the black-hole shadow in \sgra}

\author{Dimitrios Psaltis, Feryal \"Ozel, Chi-Kwan Chan, and Daniel
  P.\ Marrone}

\affil{Astronomy Department, University of Arizona, 933 N.\ Cherry Ave,
Tucson, AZ~85721}

\begin{abstract}
The half opening angle of a Kerr black-hole shadow is always equal to
$(5\pm 0.2) GM/Dc^2$, where $M$ is the mass of the black hole and $D$
is its distance from the Earth. Therefore, measuring the size of a
shadow and verifying whether it is within this 4\% range constitutes a
null hypothesis test of General Relativity. We show that the black
hole in the center of the Milky Way, \sgra, is the optimal target for
performing this test with upcoming observations using the Event
Horizon Telescope.  We use the results of optical/IR monitoring of
stellar orbits to show that the mass-to-distance ratio for \sgra is
already known to an accuracy of $\sim 4$\%. We investigate our prior
knowledge of the properties of the scattering screen between
\sgra\ and the Earth, the effects of which will need to be corrected
for in order for the black-hole shadow to appear sharp against the
background emission.  Finally, we explore an edge detection scheme for
interferometric data and a pattern matching algorithm based on the
Hough/Radon transform and demonstrate that the shadow of the black
hole at 1.3~mm can be localized, in principle, to within $\sim
9$\%. All these results suggest that our prior knowledge of the
properties of the black hole, of scattering broadening, and of the
accretion flow can only limit this General Relativistic null
hypothesis test with Event Horizon Telescope observations of \sgra\ to
$\lesssim 10$\%.
\end{abstract}


\section{INTRODUCTION}

One of the most remarkable predictions of strong-field gravity is the
existence of a region outside a black-hole horizon, in which there are
no closed photon orbits.  All photons that venture into this region
eventually cross the horizon and are removed from the observable
Universe. The net result is a shadow imprinted by every black hole on
the emission that originates in its vicinity (Bardeen 1973; Luminet
1979).

The size of the shadow of a black hole is determined by the radius of
the photon orbit in its spacetime.  The radius of the photon orbit
changes significantly with black-hole spin, from $3GM/c^2$ for a
non-spinning black hole to $GM/c^2$ for a maximally spinning one in
Boyer-Lindquist coordinates and for a prograde photon orbit (Bardeen
et al.\ 1972; here $M$ is the mass of the black hole), However,
because of gravitational lensing, the size and shape of the black-hole
shadow observed at infinity has a very weak dependence on the
black-hole spin or the orientation of the observer. Its radius changes
from $\sqrt{27}\simeq 5.2~M$ for a non-spinning black hole
(independent of orientation) to $4.84~M$ for a maximally spinning one
viewed pole on (see, e.g., de Vries 2000; Takahashi 2004; Bozza et
al.\ 2006; Johannsen \& Psaltis 2010; Chan et al.\ 2013). As a result,
for a black hole of known mass and distance, identifying the presence
of the shadow and confirming that its size is in the narrow range
$(4.8-5.2)~M$ constitutes a null hypothesis test of the Kerr
metric. Indeed, for other plausible black-hole or naked-singularity
metrics the size of the shadow has been shown to be significantly
different (for different modifications to the Kerr metric see, e.g.,
Bambi \& Freese 2009; Johannsen \& Psaltis 2010; Bambi \& Yoshida
2010; Amarilla \& Eiroa 2013).

Current 1.3~mm VLBI observations of the black hole in the center of
the Milky Way, \sgra, (Doeleman et al.\ 2008; Fish et al.\ 2009, 2014)
and earlier theoretical expectations (e.g., Falcke et al.\ 2000;
\"Ozel et al.\ 2000) all but ensure that upcoming observations with
the complete Event Horizon Telescope (EHT) will generate horizon-scale
images of its accretion flow and of the black-hole shadow (see, e.g.,
numerical simulations of Broderick et al.\ 2009, 2011; Dexter et
al.\ 2009, 2010; Mo{\'s}cibrodzka et al.\ 2009, 2013, 2014; Chan et
al.\ 2015a).  In this article, we explore the prospect of performing
the general relativistic null hypothesis test discussed above using
the upcoming EHT observations of \sgra. In this initial study, we
investigate the various systematic uncertainties that arise from our
incomplete prior knowledge of the properties of the black-hole, of the
intervening medium between the Galactic Center and the Earth, and of
the astrophysical complexities that affect the measurement of the size
of the black-hole shadow. With these uncertainties under control, we
will then be able to assess, in future work, the scatter that is
introduced by the inherent uncertainties in the measurements, the
incomplete coverage of the interferometric $u-v$ plane by the EHT
array, and the particular algorithms for image reconstruction that
will be employed.

In \S2, we investigate our prior knowledge of the mass $M$ of
\sgra\ and its distance $D$ from Earth, the ratio of which determines
the apparent size of the black-hole shadow on the sky. We show that
currently available models of the orbits of nearby stars developed by
two groups using observations with Keck (Ghez et al.\ 2008;
  Boehle et al.\ 2015) and with the VLT (Gillessen et al.\ 2009a,
  2009b) give consistent results and constrain the ratio $M/D$ to
within $\simeq 4$\%. This accuracy will only improve with next
generation adaptive optics systems on 30-m class telescopes (Weinberg
et al.\ 2005), with interferometric observations of stellar orbits
(Eisenhauer et al.\ 2011), or with the discovery of a radio pulsar in
orbit around the black hole (Pfahl \& Loeb 2004; Liu et al.\ 2012;
Psaltis et al.\ 2015b).

In \S3, we study the inferred properties of the scattering screen
between the Earth and the Galactic Center that blurs the image of the
shadow. As demonstrated in Fish et al.\ (2014), blurring due to
scattering takes place in the ensemble average regime (see Narayan \&
Goodman 1989; Goodman \& Narayan 1989). As a result, the blurring
effects on the image can be formally corrected for, if the scattering
kernel at 1.3~mm is known {\em a priori\/}. After collecting all the
published measurements of the scattered broadened image of \sgra, we
find that different aspects of the scattering ellipse at longer
wavelengths can be inferred with a fractional accuracy in the range
$3-20$\%. We argue that extrapolating these results down to the 1.3~mm
wavelength of the EHT observations requires a more accurate
measurement of the long-wavelength scattering kernel as well as a
better understanding of the wavelength dependence of its anisotropy.

Finally, in \S4 we investigate model independent methods to measure
the shape and size of the black-hole shadow from interferometric
observations. We explore an edge detection algorithm based on the
gradient of the image brightness that allows us to extract the shape
of the shadow from the image and a pattern matching algorithm based on
the Hough/Radon transform to measure the shadow properties. We apply
these algorithms to our recent ray tracing calculation of GRMHD
simulations and find that, in principle, the shape and size of the
black-hole shadow can be measured to a $\sim 9$\% accuracy.

\section{THE APPARENT SIZE OF THE BLACK-HOLE SHADOW}

General relativity predicts the size of a black-hole shadow in units
of the gravitational radius $GM/c^2$, which we will denote hereafter
simply as $M$. However, in order to compare the theoretical
predictions to the observations, we need a prior knowledge of the
ratio $M/D$ of the mass of the black hole to its distance from Earth.
This ratio sets the angular size of one gravitational radius on the
sky of an observer.

The mass of \sgra\ and its distance from the Earth can be inferred by
fitting Keplerian orbits to optical/IR observations of stars in the
vicinity of the black hole (see, e.g., Ghez et al.\ 2008; Gillessen et
al.\ 2009a, 2009b). These measurement are typically highly
correlated. Astrometric observations of the stellar positions
typically constrain the ratio $M/D^3$, whereas spectral measurements
of radial velocities constrain the ratio $M/D$. When multiple such
data sets are combined, the correlated errors between mass and
distance stretch along the curve $M\sim D^2$.

\begin{figure}[t]
\psfig{file=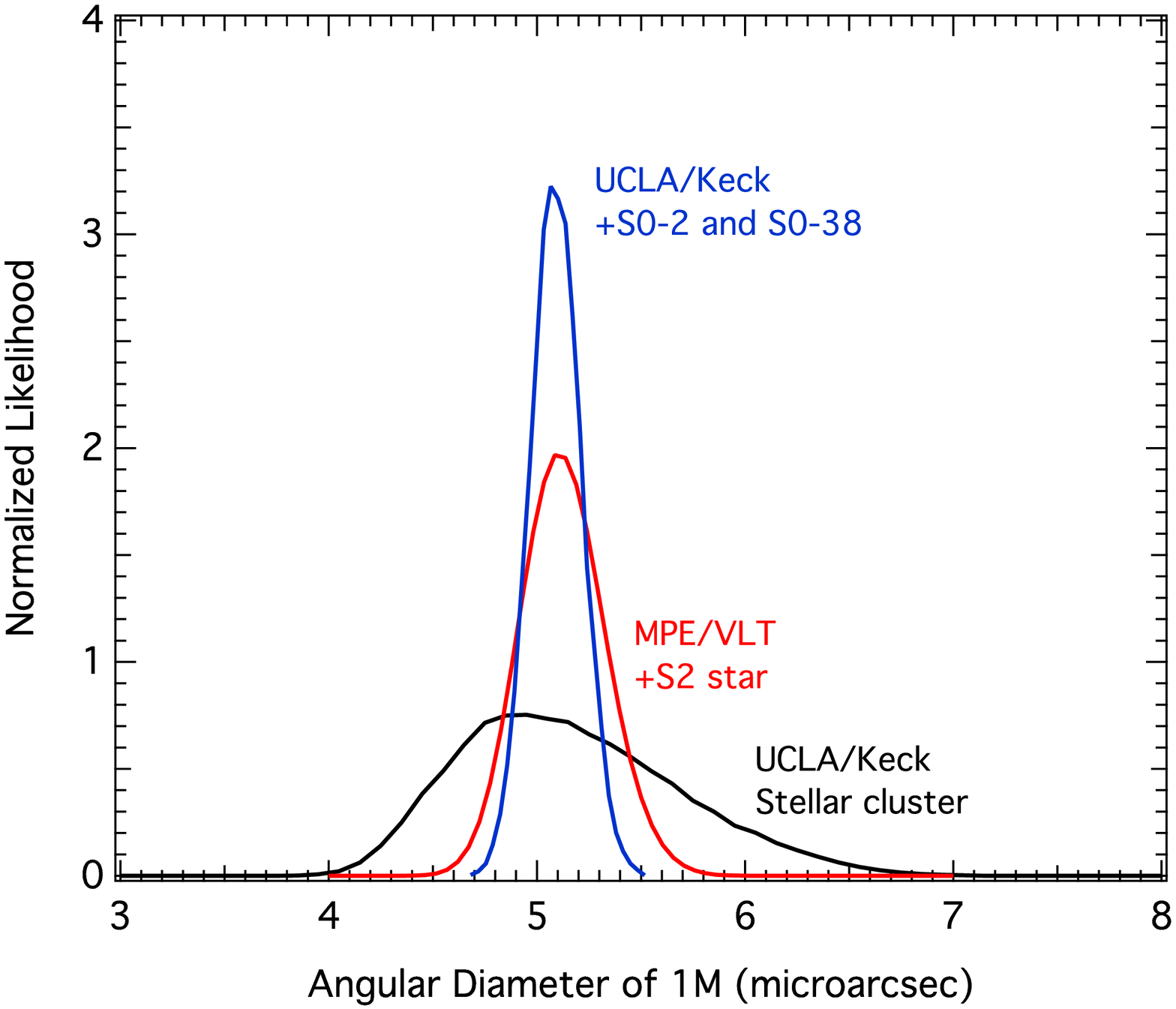,width=3.5in,height=3.0in}
\caption{The posterior likelihood of the angular size of one
  gravitational radius (i.e., $GM/c^2D$) for \sgra, as inferred from
  fitting Keplerian orbits to astrometric observations of S stars. The
  black curve shows the result of a UCLA/Keck study, in which only the
  astrometric observations of the inner stellar cluster were
  considered (Ghez et al.\ 2008). The red curve shows the result of an
  MPE/VLT study, in which the radial velocities of the S2 star were
  also incorporated (Gillessen et al.\ 2009a, 2009b). The blue curves
  shows the result of the most recent UCLA/Keck study, which also
  employs data for the S0-38 star (Boehle et al.\ 2015). The posterior
  likelihood in the Boehle et al.\ (2015) analysis corresponds to an
  angular size of one gravitational radius that is equal to 5.09$\pm
  0.17~\mu$arcsec.}
\label{fig:angdiam}
\end{figure}

In general, each study of the stellar orbits around \sgra\ leads to
the two-dimensional posterior likelihood $P(D,M) dD dM$.  For our
purposes, we are only interested in the marginalized posterior
likelihood over the ratio $P(M/D)d(M/D)$. If we set $\theta\equiv M/D$
such that
\begin{equation}
\left\vert\frac{d\theta}{dD}\right\vert=\frac{M}{D^2}
\end{equation}
and
\begin{equation}
P(\theta,M)d\theta dM=P(D,M)dD dM
\end{equation}
then we conclude that
\begin{equation}
P(\theta)d\theta=\int_M P(\theta,M)dMd\theta = \int_M \frac{D^2}{M}
P(D,M) dM d\theta\;,
\end{equation}
Using the two-dimensional posterior likelihood $P(D,M)$ over the
  distance and the mass of the black hole from studies of stellar
  orbits, we can employ this equation to calculate the posterior
  likelihood over the angular size of one gravitational radius for
  \sgra.

Figure~\ref{fig:angdiam} shows the improvement in the inference
  of this angular size, as increasingly more data are incorporated for
  the stellar orbits. When only the astrometric orbits of the inner
  stellar cluster are considered (i.e., using the two-dimensional
  posterior likelihoods $P(D,M)$ shown in Figure~11 of Ghez et
  al.\ 2008), the uncertainty in the angular size of \sgra\ is at the
  10\% level. When radial velocity data for the S2 star are also
  incorporated (i.e., using the equivalent of Figure~15 of Gillessen
  et al.\ 2009b but including the data for the S2 orbit, as discussed
  in Gillessen et al.\ 2009a), the angular size of one gravitational
  radius for \sgra\ becomes 5.12$\pm$0.29~$\mu$arcsec.  Finally, when
  the data for the S0-38 orbit are also included (Boehle et
  al.\ 2015), then the angular size of one gravitational radius
  becomres equal to 5.09$\pm$0.17~$\mu$arcsec. This $\sim 3-4$\%
uncertainty in the last measurement is comparable to the $\pm 4$\%
uncertainty in the predicted size of the black-hole shadow given that
we currently do not have a prior knowledge of the spin of the black
hole or of the orientation of its spin axis with respect to the Earth.

A number of different observations in the near future will reduce
significantly the uncertainties in the above measurements. Next
generation adaptive optics systems operating in 30-m class telescopes
may discover stars with orbits that are closer to the black hole than
the currently known S stars and that display different relativistic
effects.  The improved accuracy of the astrometric measurements and
the detection of relativistic precession will lead to uncertainties in
the inferred mass of \sgra\ and its distance that are as small as
0.1\% (Weinberg et al.\ 2005). At the same time, the detection of even
a single radio pulsar in a sub-year orbit around \sgra\ will lead to
an unprecedentent fractional accuracy in the measurement of the
black-hole mass and distance that may be as low as $10^{-6}$ (Liu et
al.\ 2012; see also Pfahl \& Loeb 2004; Psaltis et al.\ 2015b).

\section{THE SCATTERING SCREEN}

\begin{figure}[t]
\psfig{file=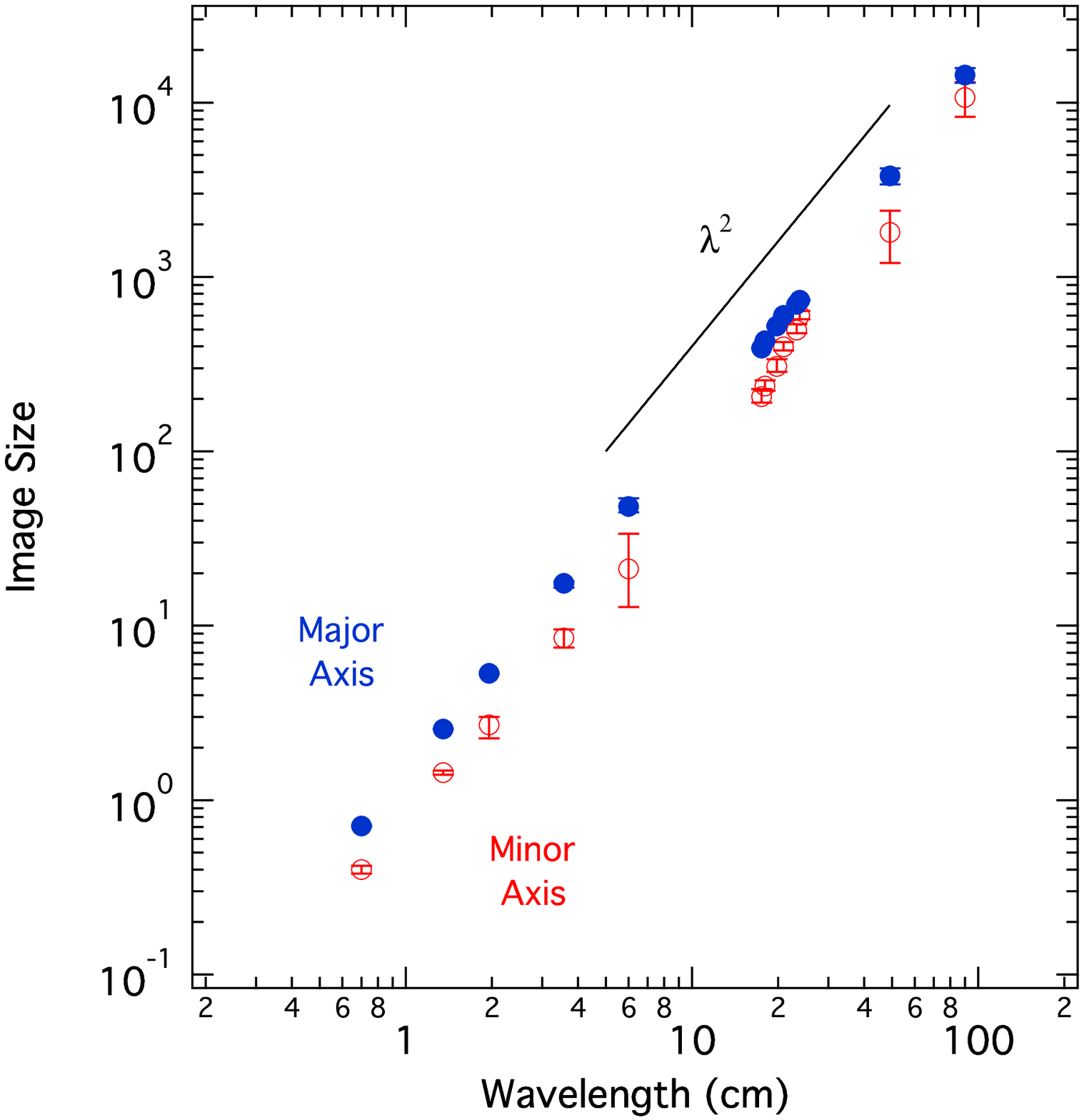,width=3.5in,height=3.0in}
\caption{The observed dependence on wavelength of the major and minor
  axis dimensions of the scatter broadened image of \sgra. The solid
  black line shows the expected $\lambda^2$ dependence.}
\label{fig:images}
\end{figure}

The long-wavelength images of \sgra\ are much larger than expected for
the emission from the accretion flow and their size depends on
wavelength in a manner that is consistent with their being
scatter-broadened by the interstellar medium (see
Figure~\ref{fig:images}, Bower et al.\ 2006, and references below). At
the 1.3~mm wavelength of the EHT observations, the size of the
scattering Kernel is expected to be of the order of 22~$\mu$arcsec,
which is approximately one fourth of the size of the black-hole
shadow. The effect of scattering is to smooth the underlying image
structure at scales smaller than the size of the Kernel, potentially
masking and blurring the edges of the shadow.

Fish et al.\ (2014) argued that scatter broadening towards
\sgra\ takes place in the ensemble average regime (Narayan \& Goodman
1989; Goodman \& Narayan 1989) and is, therefore, deterministic.
Recent observations of refractive substructure at larger wavelengths
(Gwinn et al.\ 2014), however, clearly suggest that the averaging over
the scattering ensemble is incomplete and a subsequent theoretical
study (Johnson \& Gwinn 2015) demonstrated that this substructure is
not quenched by the presence of an extended source. Nevertheless, if
data from many epochs and hence from many realizations of the
scattering ensemble are added, the effect of scatter broadening can be
considered as taking place in the ensemble average regime.

Formally speaking, scatter broadening in this regime
causes the amplitudes of the observed interferometric visibilities to
be reduced by a real factor that is equal to the Fourier transform of
the scattering kernel. If we have good prior knowledge of the
scattering kernel, we can then easily correct for the blurring effects
of scattering. Fish et al.\ (2014) explored different avenues of
deblurring (direct or with Wiener filters) and demonstrated their
efficiency with mock EHT data. They also considered the formal
uncertainties in the measurement of the scattering kernel properties
(as reported by Bower et al.\ 2006) and their effect on the results
of the deblurring algorithms.

In this section, we explore the systematic uncertainties in the
measurements of the properties of the scattering kernel. This is
important for a number of reasons, which we discuss below.

Measurements of the scattering kernel properties need to be done at
relatively long wavelengths, for which the size of the source is much
smaller than the size of the kernel. Recent observations found
evidence for milliarcsec substructure in the images of \sgra\ at
wavelengths as large as 1.3~cm (Gwinn et al.\ 2014). This is
corroborated by the measured deviation of the wavelength dependence of
the scattering kernel from the theoretical expected scaling of
$\lambda^2$ at comparable wavelengths, as we show in
Figure~\ref{fig:normalized} below (see also Falcke \&
Markoff~2012). As a result, the properties of the scattering kernel
need to be measured at wavelengths as large as $10-20$~cm and then
extrapolated down to the 1.3~mm wavelength of the EHT. This
extrapolation by two orders of magnitude runs the risk of amplifying
even minor systematic errors that are unaccounted for.

The large dynamical range of the extrapolation becomes even more
important when we consider the fact that the measured scattering
kernel towards \sgra\ is highly anisotropic. It is often modeled in
terms of an ellipse, with a 2:1 axis ratio and a major axis that is
oriented primarily along the E-W axis on the sky (see \S3.1
below). The origin of the anisotropy is widely believed to be related
to large scale magnetic fields in the turbulent interstellar medium,
the statistical properties of which determine both the axis ratio of
the scattering ellipse and its orientation (see, e.g., Chandran \&
Backer 2002). These two quantities may not remain constant, as is
currently assumed, when the wavelength changes by over two orders of
magnitude. A theoretical investigation of this issue is necessary to
make progress but is beyond the scope of the present work.

Our goal in this section is to explore the wavelength dependence of
the properties of the scattering kernel towards \sgra\ and infer
the level of existing systematic differences between the various 
measurements across the mm-to-cm spectrum. 

\subsection{Observational Constraints}

Tables~\ref{tab:var}-\ref{tab:shen} summarize the published
measurements of the image size of \sgra\ in the mm-to-cm range, for
which the data quality was sufficient to detect an elliptical (as
opposed to a circular) image. We only consider measurements at
wavelengths larger than 0.7~mm, as source structure is clearly seen at
1.3 and 3~mm (Doeleman et al.\ 200; Bower et al.\ 2014). For most
wavelengths, there has been only one measurement reported, which is
what we include. For some wavelengths, more than one measurements have
been carried out; in these cases, as we explain below, we chose the
result of the most recent publication (when the same data were
analyzed more than once) or the one with the smallest reported
uncertainties (which is often the most recent, as well).

\begin{figure*}[h]
\psfig{file=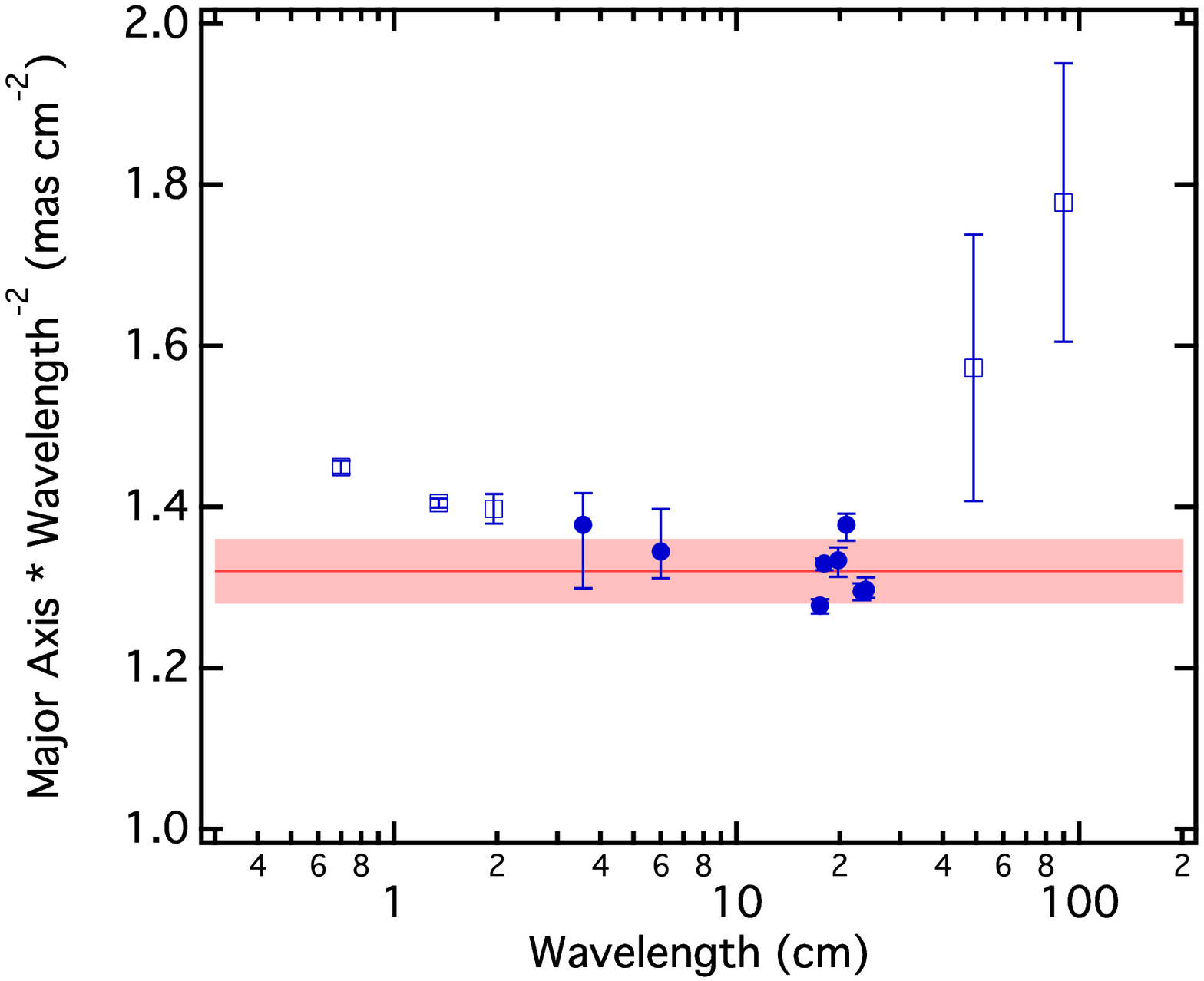,width=3.5in}
\psfig{file=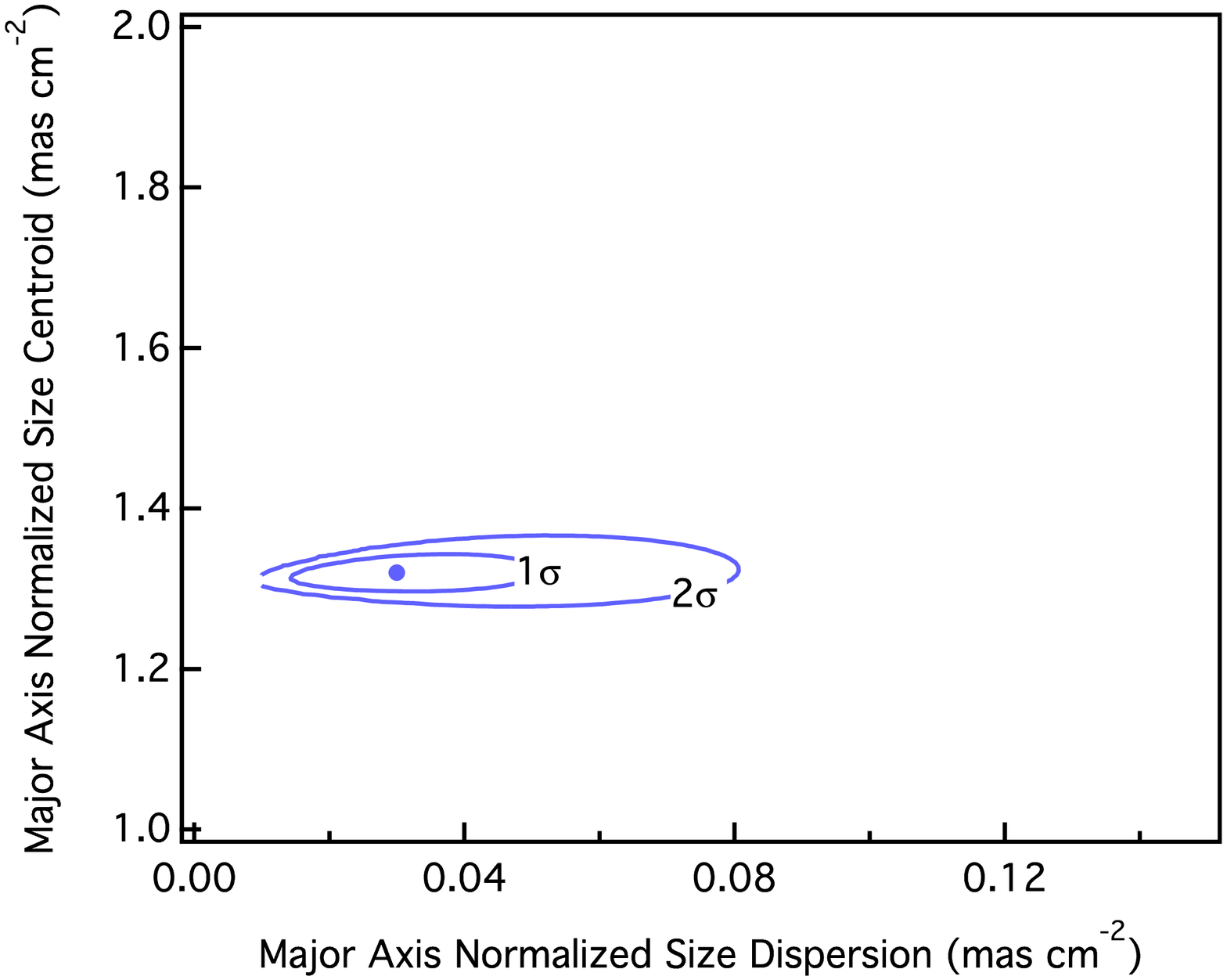,width=3.5in}
\psfig{file=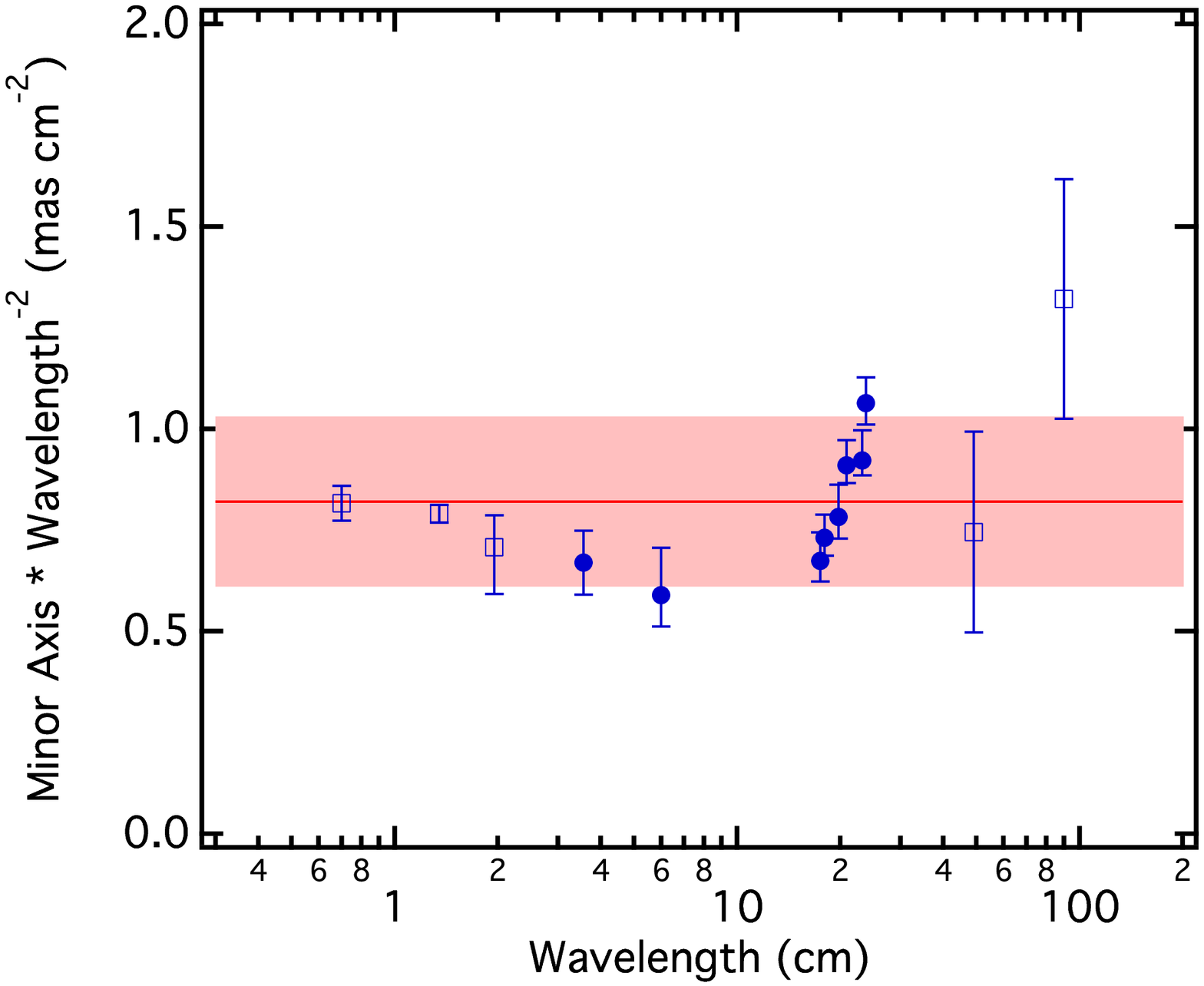,width=3.5in}
\psfig{file=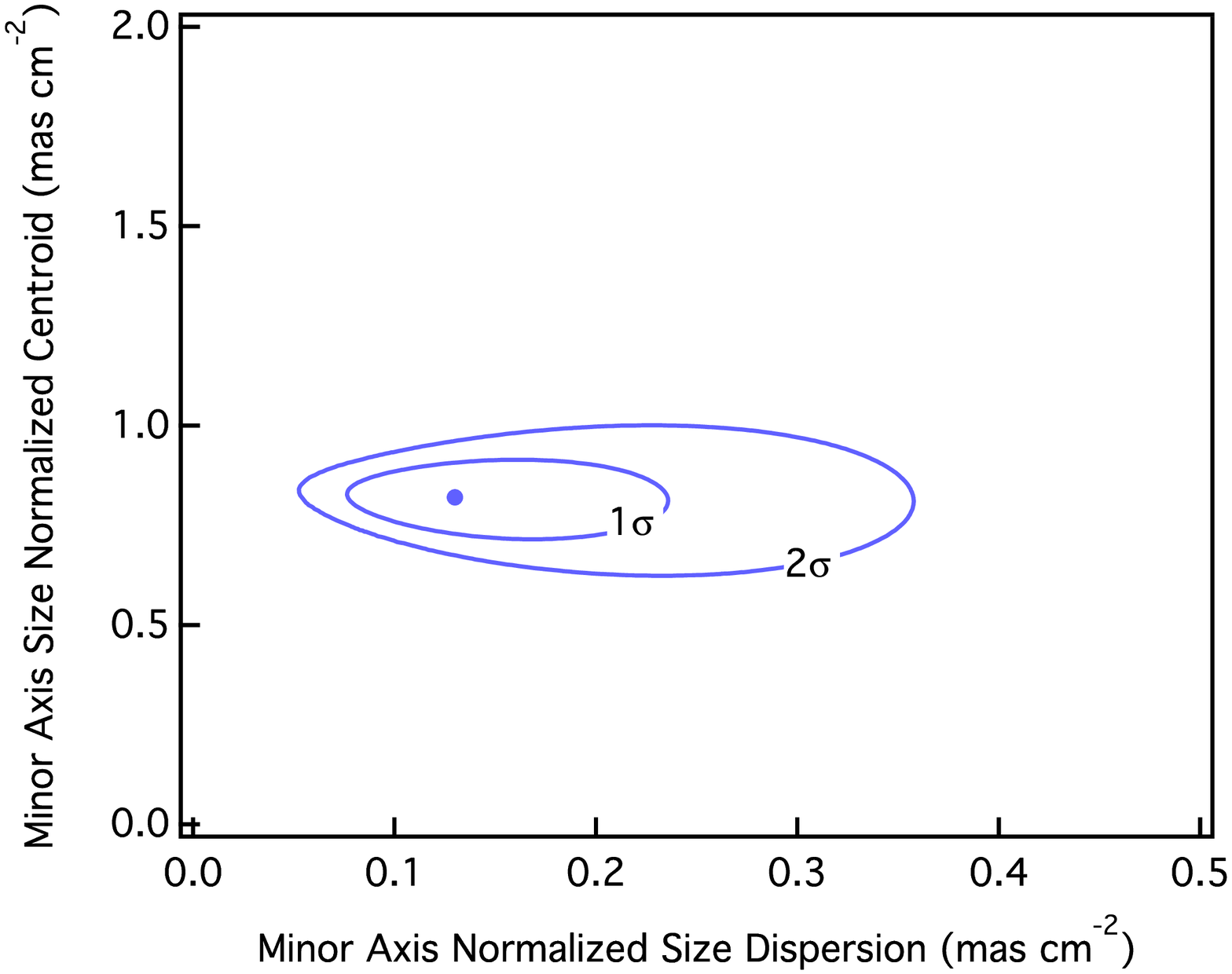,width=3.5in}
\psfig{file=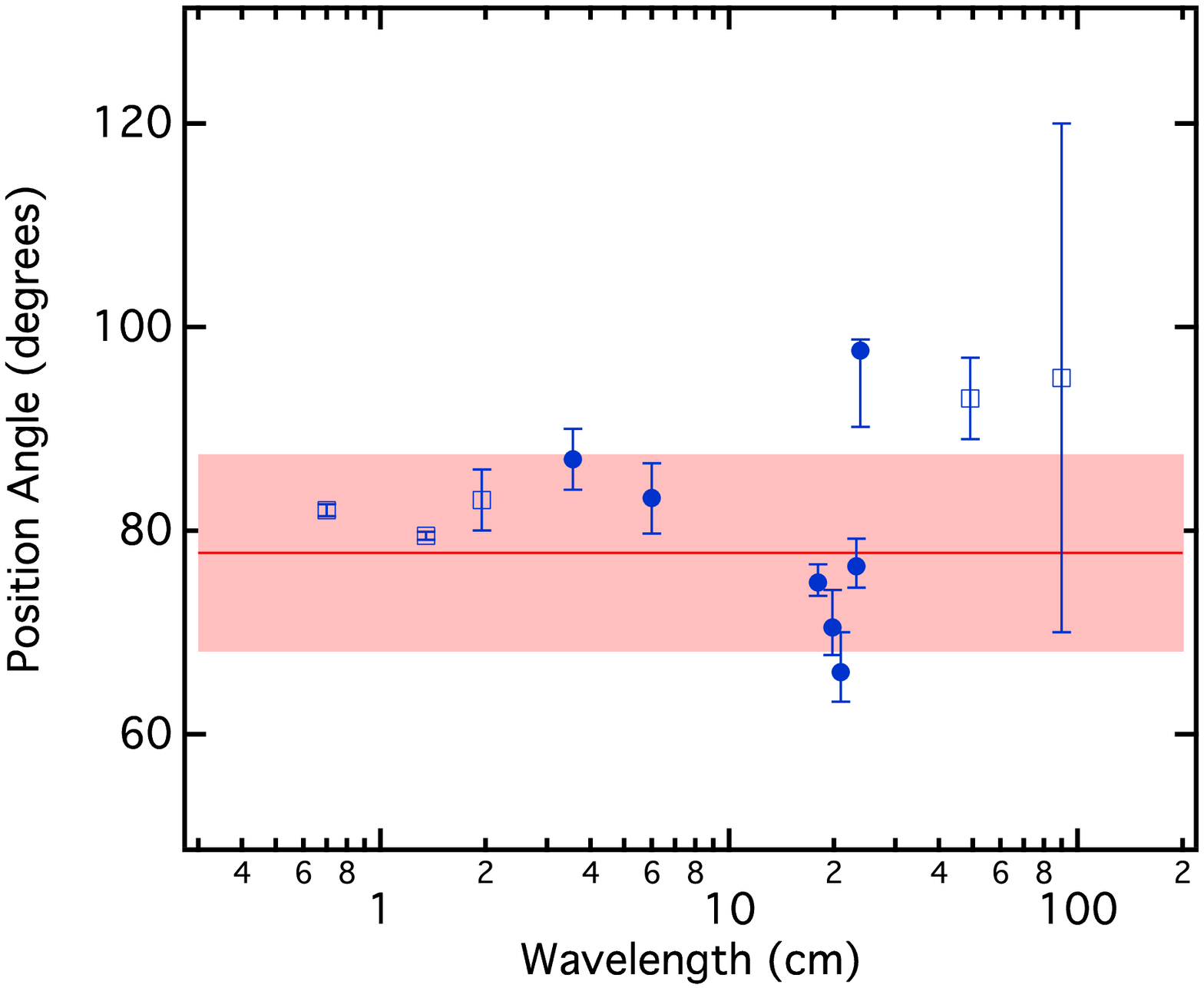,width=3.5in}
\psfig{file=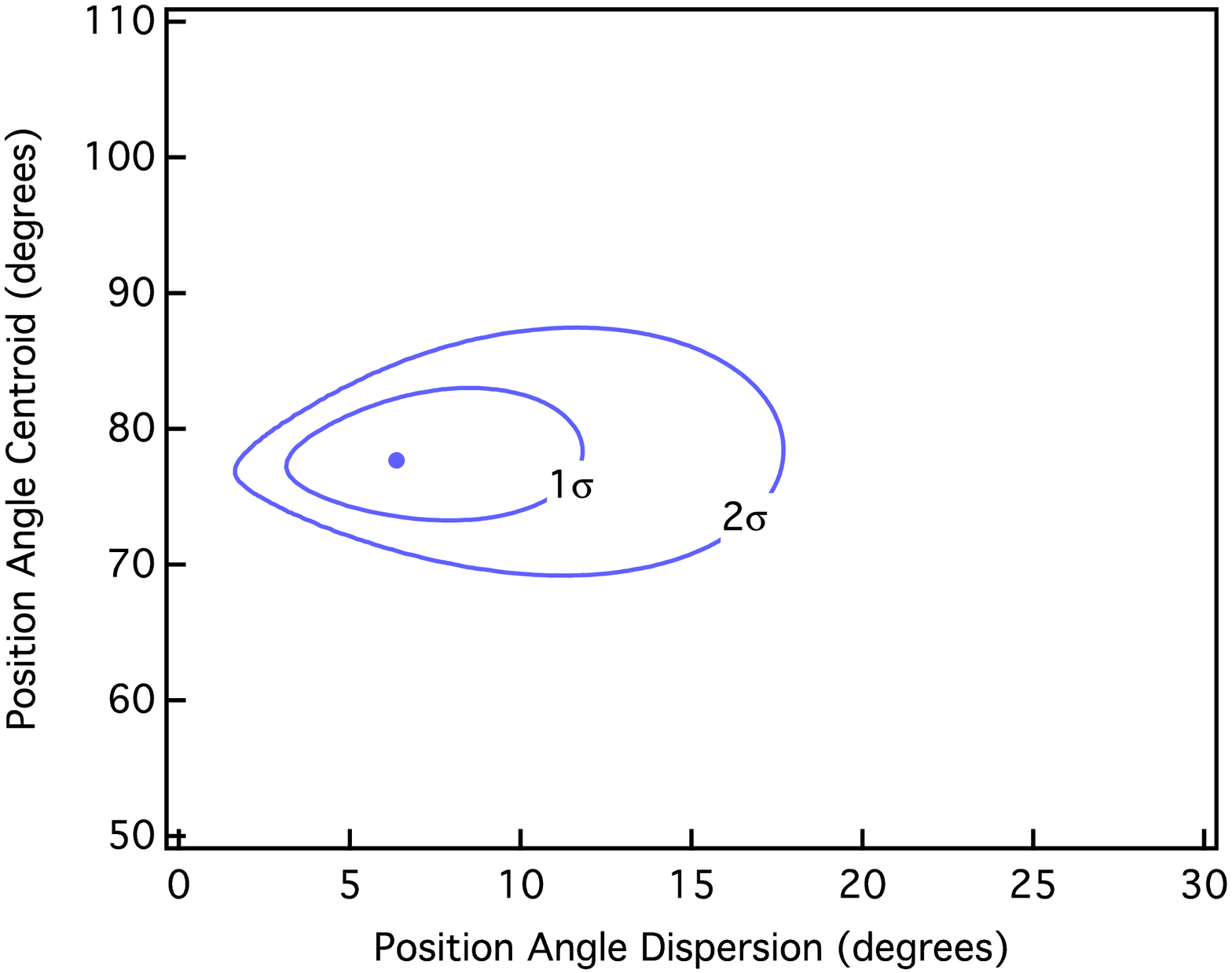,width=3.5in}
\caption{{\em (Left)\/} The ratio of the measured {\em (Top)\/} major
  axis and {\em (Middle)\/} minor axis to the square of the
  wavelength, as a function of the observed wavelength. {\em (Bottom)}
  The position angle of the scattering ellipse, as a function of the
  observed wavelength. In each panel, the filled circles denote the
  data points that were used in inferring the average properties of
  the scattering screen; the colored bands show the 68\% posterior
  likelihood range of parameters of the scattering screen.  {\em
    (Right)\/} The 2-dimensional posterior likelihoods of the three
  parameters shown on the left panels. In all three cases, the
  systematic uncertainties are assumed to be described by a Gaussian
  distribution with a given centroid (plotted along the y-axis) and
  dispersion (plotted along the x-axis). The most likely value of the
  centroid and dispersion are denoted by a filled circle in each
  panel, whereas the 68\% and 95\% contours are also shown as
  continuous lines.  }
\label{fig:normalized}
\end{figure*}

\begin{deluxetable*}{ccccc}
  \tablecolumns{5}
  \tablewidth{400pt}
\tablecaption{Elliptical Image Sizes of \sgra}
  \tablehead{\colhead{Wavelength} & \colhead{Major Axis}  & 
\colhead{Minor Axis}  & \colhead{P.A.\tablenotemark{a}} & \colhead{Reference}\\
 \colhead{(cm)} & \colhead{(marcsec)} & \colhead{(marcsec)} & \colhead{(degrees)} & }
\startdata
90	& 14400$\pm$1400 & 10700$\pm$2400 & 95$\pm$25  &An et al. 2005\\
49.16	& 3800$\pm$400   & 1800$\pm$600	  & 93$\pm$4   &Roy \& Pramesh Rao 2003\\
23.8	& 734.7$^{+8.7}_{-5.9}$  & 602.4$^{+36.0}_{-29.9}$ & 97.7$^{+1.1}_{-7.5}$ & Bower et al.\ 2006\\ 
23.2	& 697.1$^{+5.4}_{-5.9}$  & 496.3$^{+40.1}_{-19.6}$ & 76.5$^{+2.7}_{-2.1}$ & Bower et al.\ 2006\\
20.9	& 601.8$^{+6.0}_{-8.6}$  & 397.3$^{+27.2}_{-19.1}$ & 66.1$^{+3.9}_{-2.9}$ & Bower et al.\ 2006\\
19.8	& 522.8$^{+6.3}_{-8.0}$  & 306.6$^{+31.2}_{-21.1}$ & 70.5$^{+3.7}_{-2.7}$ & Bower et al.\ 2006\\
18	& 430.8$^{+1.9}_{-2.8}$  & 236.7$^{+18.6}_{-14.4}$ & 74.9$^{+1.8}_{-1.3}$ & Bower et al.\ 2006\\
17.5	& 391.2$^{+2.4}_{-3.0}$  & 206.3$^{+21.6}_{-15.5}$ & 74.5$^{+2.1}_{-1.5}$ & Bower et al.\ 2006\\
6	& 48.4$^{+1.9}_{-1.2}$   & 21.2$^{+4.2}_{-2.8}$   & 83.2$^{+3,4}_{-3.5}$ & Bower et al.\ 2004\tablenotemark{b}\\
3.56    & 17.5$^{+0.5}_{-1.0}$   & 8.5$^{+1.0}_{-1.0}$     & 87$^{+3}_{-3}$      & Shen et al.\ 2005\\
1.95    & 5.33$^{+0.07}_{-0.07}$ & 2.7$^{+0.3}_{-0.44}$    & 83$^{+3}_{-3}$      & Shen et al.\ 2005\\
1.35	& 2.56$\pm$0.01         & 1.44$\pm$0.04         & 79.5$\pm$0.4           & Lu et al.\ 2011\\
0.7	& 0.722$\pm$0.004       & 0.40$\pm$0.02         & 82.0$\pm$0.6           & Lu et al.\ 2011
\enddata
\tablenotetext{a}{Position angle of the major axis, measured in degrees
East of North.}
\tablenotetext{b}{The uncertainties displayed here are 1/3 of the 3$\sigma$
uncertainties quoted in this reference.}
\label{tab:var}
\end{deluxetable*}

\begin{deluxetable*}{cccc}
  \tablecolumns{4}
  \tablewidth{350pt}
\tablecaption{Elliptical Image Sizes of \sgra\ reported by Lu et al.\ 2011}
  \tablehead{\colhead{Wavelength} & \colhead{Major Axis}  & 
\colhead{Minor Axis}  & \colhead{P.A.\tablenotemark{a}} \\
 \colhead{(cm)} & \colhead{(marcsec)} & \colhead{(marcsec)} & \colhead{(degrees)}}
\startdata
1.35 & 2.56$\pm$0.03  & 1.51$\pm$0.10 &  78.6$\pm$1.7\\
     & 2.55$\pm$0.06  & 1.28$\pm$0.08 &  82.0$\pm$1.1\\
     & 2.55$\pm$0.04  & 1.61$\pm$0.13 &  78.0$\pm$1.7\\
     & 2.56$\pm$0.05  & 1.54$\pm$0.08 &  79.8$\pm$1.0\\
     & 2.53$\pm$0.03  & 1.44$\pm$0.05 &  79.6$\pm$1.3\\
     & 2.53$\pm$0.02  & 1.34$\pm$0.05 &  80.2$\pm$1.3\\
     & 2.58$\pm$0.02  & 1.57$\pm$0.05 &  77.2$\pm$0.8\\
     & 2.56$\pm$0.03  & 1.48$\pm$0.05 &  78.9$\pm$0.5\\
     & 2.55$\pm$0.03  & 1.38$\pm$0.10 &  80.7$\pm$0.8\\
     & 2.57$\pm$0.01  & 1.39$\pm$0.05 &  81.1$\pm$0.9\\
0.7  & 0.71$\pm$0.01  & 0.41$\pm$0.04 &  81.7$\pm$2.3\\
     & 0.72$\pm$0.01  & 0.38$\pm$0.03 &  82.1$\pm$0.8\\
     & 0.72$\pm$0.01  & 0.45$\pm$0.04 &  82.0$\pm$2.4\\
     & 0.71$\pm$0.01  & 0.38$\pm$0.06 &  84.1$\pm$1.1\\
     & 0.71$\pm$0.01  & 0.36$\pm$0.04 &  84.7$\pm$3.1\\
     & 0.72$\pm$0.01  & 0.39$\pm$0.04 &  80.4$\pm$1.6\\
     & 0.72$\pm$0.01  & 0.44$\pm$0.04 &  81.8$\pm$3.1\\
     & 0.72$\pm$0.01  & 0.48$\pm$0.04 &  78.5$\pm$1.3\\
     & 0.72$\pm$0.01  & 0.39$\pm$0.04 &  81.2$\pm$1.3\\
     & 0.68$\pm$0.01  & 0.33$\pm$0.04 &  86.6$\pm$2.1
\enddata
\tablenotetext{a}{Position angle of the major axis, measured in degrees
East of North.}
\label{tab:lu}
\end{deluxetable*}

\begin{deluxetable*}{cccc}
  \tablecolumns{4}
  \tablewidth{350pt}
\tablecaption{Elliptical Image Sizes of \sgra\ at 7~mm reported by Shen et al.\ 2005}
  \tablehead{\colhead{Wavelength} & \colhead{Major Axis}  & 
\colhead{Minor Axis}  & \colhead{P.A.\tablenotemark{a}} \\
 \colhead{(cm)} & \colhead{(marcsec)} & \colhead{(marcsec)} & \colhead{(degrees)} }
\startdata
0.694 & 0.71$^{+0.01}_{-0.01}$    & 0.42$^{+0.05}_{-0.05}$    & 74$^{+2}_{-2}$ \\
0.695 & 0.722$^{+0.002}_{-0.002}$ & 0.395$^{+0.019}_{-0.020}$  & 80.4$^{+0.8}_{-0.8}$\\
0.695 & 0.725$^{+0.002}_{-0.002}$ & 0.372$^{+0.020}_{-0.018}$  & 80.4$^{+0.6}_{-0.9}$\\
0.695 & 0.72$^{+0.01}_{-0.01}$    & 0.39$^{+0.07}_{-0.07}$     & 78$^{+2}_{-2}$\\
0.695 & 0.72$^{+0.01}_{-0.01}$    & 0.42$^{+0.03}_{-0.03}$	   & 79$^{+1}_{-1}$\\
0.695 & 0.69$^{+0.01}_{-0.01}$    & 0.33$^{+0.04}_{-0.04}$     & 83$^{+1}_{-1}$\\
0.695 & 0.71$^{+0.01}_{-0.01}$	& 0.44$^{+0.02}_{-0.02}$     & 79$^{+1}_{-1}$
\enddata
\tablenotetext{a}{Position angle of the major axis, measured in degrees
East of North.}
\label{tab:shen}
\end{deluxetable*}

In particular: \newline
\noindent {\em (i)\/} At 20.9~cm, we chose the measurement reported in
Bower et al.\ (2006), as opposed to the one in Yusef-Zadeh et
al.\ (1994), because the latter has a very large reported uncertainty
in the measurement of the minor axis of the ellipse.\newline
\noindent {\em (ii)\/} At 3.5~cm, six measurements have been reported:
Lo et al.\ (1985), Jauncey et al.\ (1989), Lo et al.\ (1993), Bower et
al.\ (2004), Shen et al.\ (2005; this study included analysis of all
previous VLBA observations between 1994 and 2004) and Bower et
al.\ (2014); all of these results are statistically consistent with
each other. we chose the measurement reported in Shen et al.\ (2005)
as representative for this wavelength.\newline
\noindent {\em (iii)\/} At 2~cm, we chose the measurement reported in
Shen et al.\ (2005), as opposed to the one in Bower et al.\ (2004), because
of the large uncertainty in the measurement of the size of the minor axis
in the latter study.\newline
\noindent {\em (iv)\/} At 1.35~cm, five measurements have been
reported: Lo et al.\ (1993), Alberdi et al.\ (1993; which were
analyzed again by Marcaide et al.\ 1999 to correct an error), Bower et
al.\ (2004), Shen et al.\ (2005; this study included analysis of all
previous VLBA observations between 1994 and 2004), and Lu et
al.\ (2011). In the last study, 10 individual measurements were
obtained over the course of 10 days.  The results of these
measurements are listed in Table~\ref{tab:lu} and, for each quantity,
they are statistically consistent with each other. We combined
them, according to the method discussed in Appendix~A, assuming that
they are all drawn from the same underlying $\delta$-function
distribution. We obtained $2.56\pm 0.01$~marcsec, $1.44\pm
0.04$~marcsec and $79.5\pm 0.4^\circ$ as the most likely values of the
major axis, minor axis, and position angle, respectively. These values
are consistent with the earlier, less accurate results of Lo et
al.\ (1993), Marcaide et al.\ (1999), Bower et al.\ (2004) and Shen et
al.\ (2005).\newline
\noindent {\em (v)\/} At 0.7~cm, four sets of measurements have been
reported: Bower et al.\ (2004), Shen et al.\ (2005), Lu et
al.\ (2011), and Akiyama et al.\ (2013). All measurements are
statistically consistent with each other. However, since each data set
comprises several (up to 10) observations each, it remains a
possibility that, when the data from each study are statistically
combined, as described in Appendix~A, then systematic differences
between studies will be revealed. In order to explore this, we focused
on the studies of Shen et al.\ (2005) and Lu et al.\ (2011) and
combined the data assuming that, for each study, all measurements were
drawn from the same underlying $\delta$-function distribution. For the
measurements of Shen et al.\ (2005), we obtained $0.722\pm
0.002$~marcsec, $0.398\pm 0.014$~marcsec, and $80.0\pm 0.6^\circ$ for
the major axis, minor axis, and position angle, respectively. For the
measurements of Lu et al.\ (2011), we obtained
$0.713\pm0.004$~marcsec, $0.399\pm 0.021$~marcsec, and $82.0\pm
0.6^\circ$ for the same quantities. Both sets of measurements are
statistically consistent with each other, with similar
uncertainties. Even though we will not be using measurements at this
wavelength to quantify the degree of scattering in the image of \sgra,
we will display in the following figures the results obtained from the
measurements of Lu et al.\ (2011).

Of the various measurements included in Table~\ref{tab:var}, those of
Bower et al.\ (2004, 2006) were performed using closure amplitude
techniques, whereas the remaining followed the standard
self-calibration approach. The latter was criticized by Bower et
al.\ (2014) as being heavily dependent on the input model,
underestimating true errors. This might be explain the fact that the
various measurements are often not statistically consistent with each
other.  However, in our inference of the systematic uncertainties in
our prior knowledge of the scattering screen, we use predominantly the
measurements performed using closure amplitude techniques.

Figure~\ref{fig:images} shows the observed dependence on wavelength of
both the major and minor axis dimensions of the scattering ellipse
towards \sgra. As expected, they both show roughly a $\lambda^2$
scaling. In order to quantify this behavior and infer the level of
systematic uncertainties in the measurements, the left panels in
Figure~\ref{fig:normalized} show the ratio of the measured major and
minor axes of the scattering ellipse to the square of the wavelength
of each observation as well as the position angle of the ellipse as a
function of wavelength. It is clear from these figures that the
various measurements are not statistically consistent with each other,
with a level of systematic uncertainty that is larger than the
reported formal uncertainties.

It is well understood that, at short wavelengths, the deviation of the
scattering law from the expected $\lambda^2$ dependence is caused by
the fact that the intrinsic size of \sgra\ is comparable to or larger
than that of the scattering ellipse. On the other hand, at the largest
wavelengths, the size of the scattering ellipse is so large that the
image of \sgra\ is blurred together with that of the surrounding
emission that has significant structure.

The first goal of this exercise is to identify a dynamical range of
wavelengths in which the ratio of the major or minor axis to the
square of wavelength remains approximately
constant. Figure~\ref{fig:normalized} demonstrates that finding such a
range is rather non trivial. In order to make progress, we chose to
use the measurements with wavelengths in the range
3~cm$<\lambda<$30~cm and assign any deviation from the expected
$\lambda^2$ dependence to the budget of systematic errors.

We combined the data in this wavelength range using the procedure
discussed in Appendix~A. We assumed that all the measurements of each
property (i.e., the major axis, the minor axis, and the position
angle) were drawn from a Gaussian distribution, the widths of which
quantifies the degree of systematic uncertainties. We then obtained
the posterior likelihoods over the centroid and dispersion of each
Gaussian, which allow us to quantify the most likely value and the
systematic errors in each measurement. The results are shown in the
right panels of Figure~\ref{fig:normalized}. It is important to
emphasize here that the sizes of the two axes and the position angles
of the ellipse are not independent measurements. However, due to
absence of additional information that would allow us to quantify the
degree of correlation between the two measurements, we assumed that
they are independent in this statistical study (Bower et al.\ 2004 and
Bower et al.\ 2014 performed a detailed analysis of the correlated
uncertainties for some of the wavelengths used here).

As expected from the visual inspection of the left panels of
Figure~\ref{fig:normalized}, the inferred dispersions of the three
Gaussians are not consistent with zero, indicating the presence of
significant systematic uncertainties.  Marginalizing over the
two-dimensional likelihoods shown in the right panels of
Figure~\ref{fig:normalized}, as discussed in Appendix~A, leads to
the following measurement of the properties of the scattering ellipse
towards \sgra:
\begin{eqnarray}
\frac{\mbox{major axis size}}{\lambda^2}&=&1.32\pm 0.04~
\frac{\mbox{marcsec}}{\mbox{cm}^2}\nonumber\\
\frac{\mbox{minor axis size}}{\lambda^2}&=&0.82\pm 0.21~
\frac{\mbox{marcsec}}{\mbox{cm}^2}\nonumber\\
\mbox{P.A.}&=&77.8\pm 9.7~\mbox{degrees}\;.
\end{eqnarray}

The fractional uncertainties in the inferred size of the major axis of
the scattering ellipse is only $\simeq 3$\%. However, the fractional
uncertainties in the minor axis size and in the position angle are
$\sim 25$\% and $\sim 12$\%, respectively. The primary reason for
these large uncertainties is the fact that most measurements were done
with interferometric arrays that have a good coverage along the E-W
orientation but a poor coverage perpendicular to it. Additional
observations with better coverage in the $N-S$ orientation and a
theoretical understanding of the wavelength dependence of the axis
ratio and the position angle of the scattering ellipse will be crucial
in reducing the level of these systematic uncertainties.

\section{MEASURING THE PROPERTIES OF THE BLACK-HOLE SHADOW}

\begin{figure*}[h]
\centerline{\psfig{file=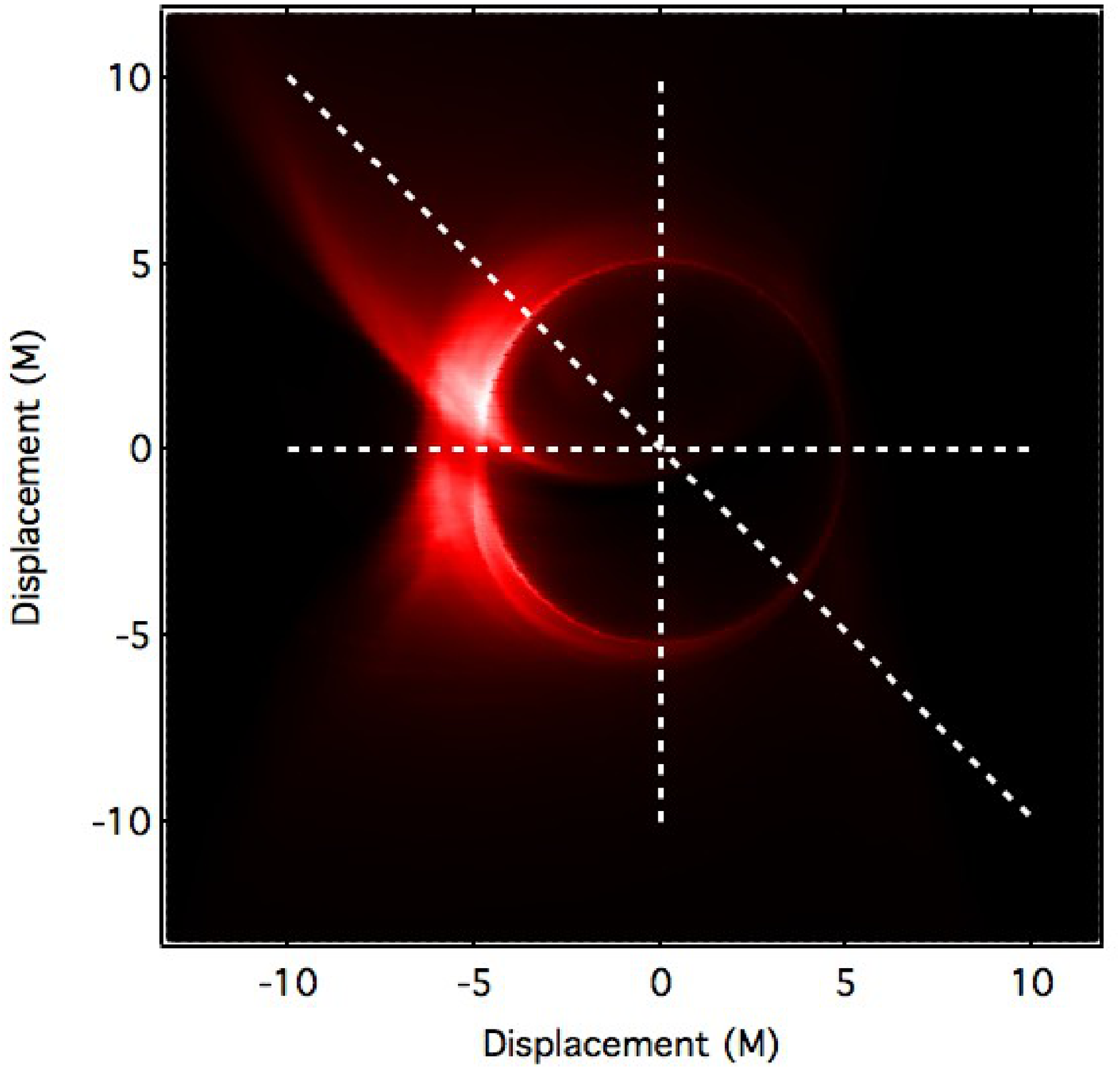,width=3.5in,height=3.0in}
\psfig{file=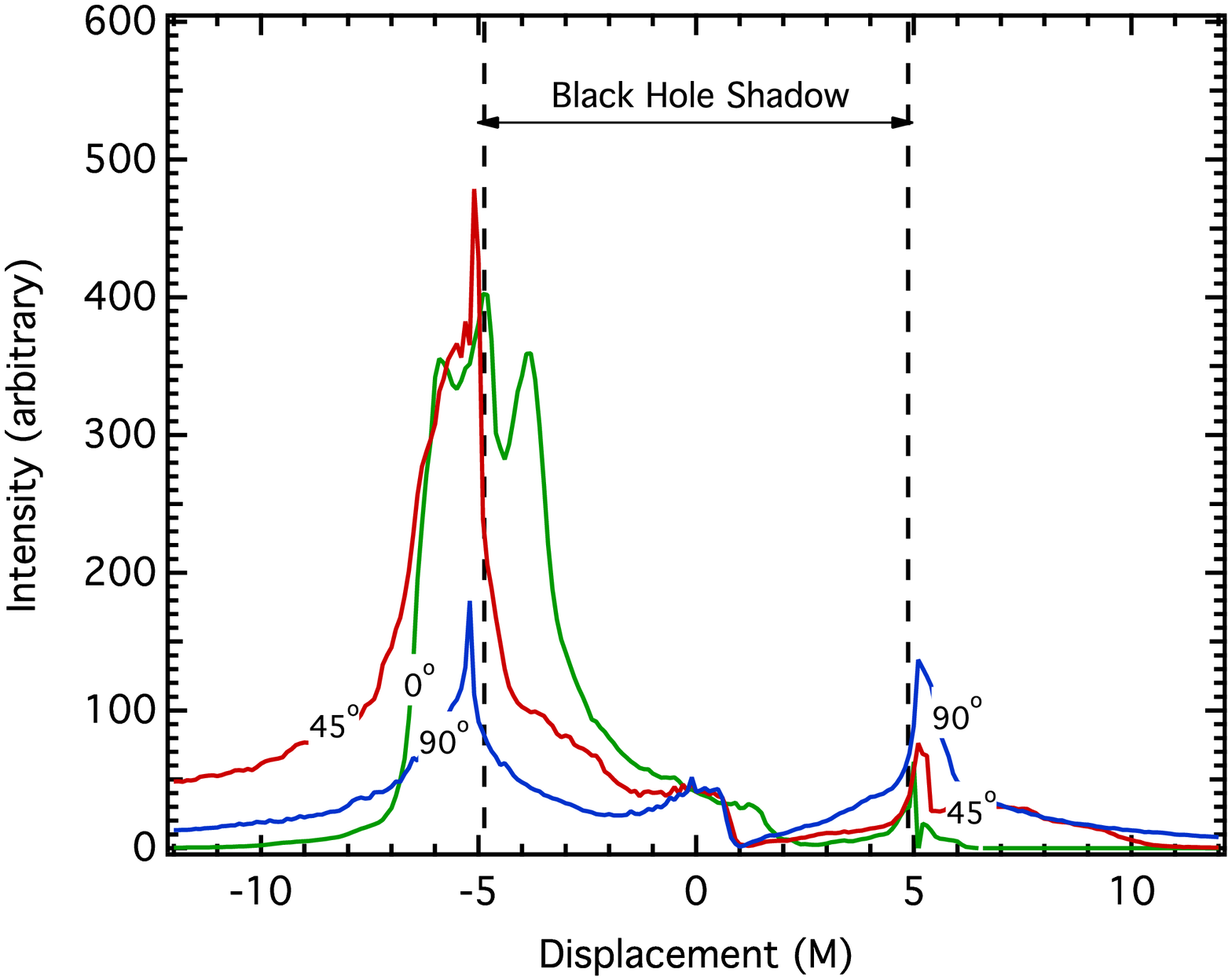,width=3.5in,height=3.0in}}
\caption{{\em (Left)\/} A simulated 1.3~mm image of \sgra, as
  calculated from a GRMHD simulation of accretion onto a black hole
  with spin $a=0.9$ (\texttt{a9SANE} simulation of Chan et
  al.\ 2015a). The inclination of the observer is set to 60$^\circ$
  with respect to the angular momentum vector of the accretion flow
  and the thermodynamic parameters of the flow were chosen such that
  the simulated spectrum is consistent with observations. The three
  dotted lines show three cross sections at 0$^\circ$, 45$^\circ$, and
  90$^\circ$ with respect to the equatorial plane. {\em (Right)\/} The
  brightness of the image shown on the left panel, along the three
  indicated cross sections.  In all cases, the rim of the black-hole
  shadow corresponds to a sharp drop in the brightness. Along most
  cross sections, the brightness drops from its maximum value at
  $\sim+0.5$~M outside the nominal location of the shadow to its
  minimum at $\sim -0.5$~M inside it. Only along the near-equatorial
  cross sections towards the Doppler boosted part of the emission
  (center left on the left image), is the transition across the shadow
  significantly smoother because of the larger emissivity of the
  intervening accretion flow.}
\label{fig:contrast}
\end{figure*}

\begin{figure*}[h]
\centerline{\psfig{file=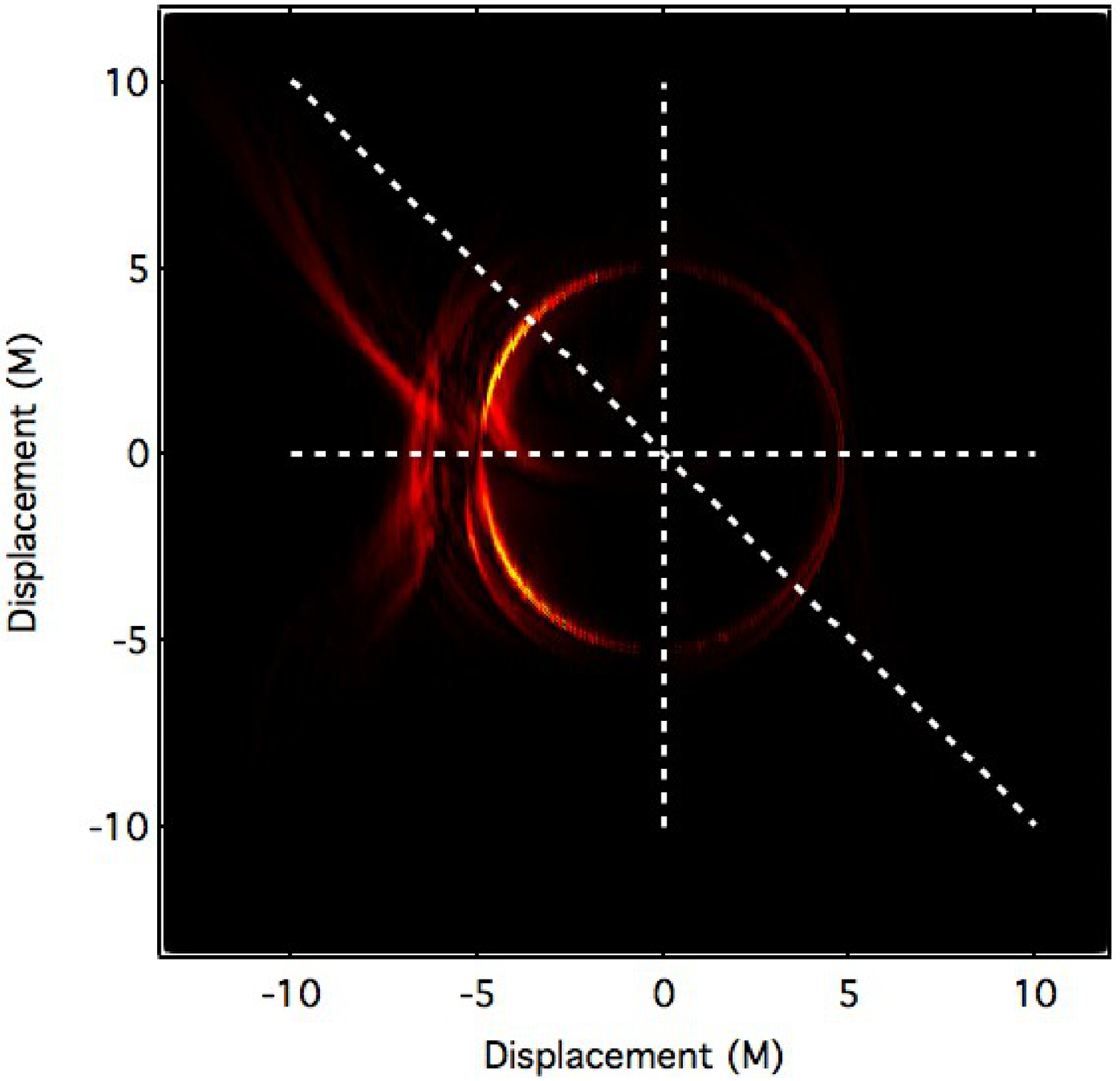,width=3.5in,height=3.0in}
\psfig{file=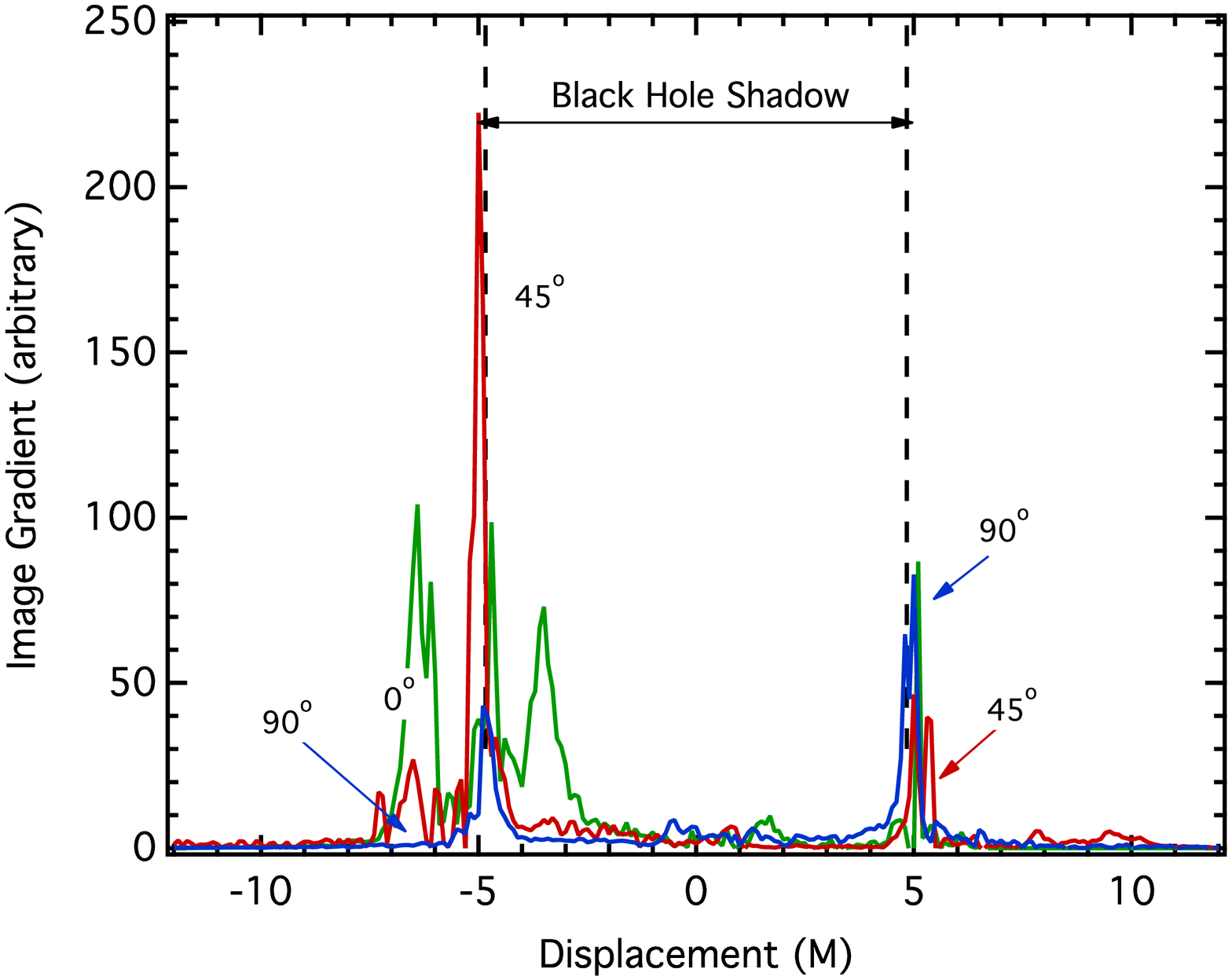,width=3.5in,height=3.0in}}
\caption{{\em (Left)\/} A map of the magnitude of the gradient of the
  image brightness shown in Figure~\ref{fig:contrast}. The bright rim
  at the location of the black-hole shadow is clearly visible. The
  three dotted lines show three cross sections at 0$^\circ$,
  45$^\circ$, and 90$^\circ$ with respect to the equatorial
  plane. {\em (Right)\/} The magnitude of the gradient of the image
  brightness shown on the left panel, along the three indicated cross
  sections. In almost all cross sections towards the Doppler boosted
  part of the emission (center left on the left image), the gradient
  shows a prominent peak within $\sim 0.25$~M of the location of the
  black-hole shadow. In the opposite side of the image, the gradients
  are more sharply peaked at displacements that are indistinguishable
  from the location of the black-hole shadow.}
\label{fig:gradients}
\end{figure*}

In its original conception, the black-hole shadow was calculated under
the assumption of back illumination of a black hole by a bundle of
photons in parallel trajectories that originated in the far infinity
(Bardeen 1973). In this case, the trajectories of photons that enter
inside the photon radius are gravitationally bent towards the
black-hole horizon and never reach the distant observer.

In the realistic case of a mm image of an accreting black hole, such
as \sgra, the situation is, of course, different, because the majority
of photons originate very close to the location of the photon orbit,
in all directions. Nevertheless, if the emission is optically thin, as
it is expected to be at mm wavelengths for \sgra, simulations have
repeatedly shown that a shadow will be imprinted on the image of the
accretion flow, (see, e.g., Jaroszynski \& Kurpiewski 1997; Falcke et
al.\ 2000; Broderick et al.\ 2009, 2011; Dexter et al.\ 2009, 2010;
Mo\'scibrodzka et al.\ 2009, 2013, 2014; Chan et al.\ 2015a). Within
the shadow, the brightness of the image is suppressed by at least an
order of magnitude, while the outline of the shadow is surrounded by a
bright photon ring (see Figure~\ref{fig:contrast}).

In principle, comparing a detailed model for the emission properties
of the accretion flow around \sgra\ to EHT observations will lead to a
measurement of the size and shape of the shadow imprinted by the black
hole. However, predicting the details of the brightness profile of the
image of the accretion flow that surrounds the black-hole shadow
beyond its gross properties is not feasible for two reasons.

The first reason is related to our current understanding of accretion
physics. In all available analytical and numerical models of the
\sgra\ emission, the thermodynamic properties of the electrons are
incorporated in a highly simplified and {\em ad hoc\/} fashion (see,
e.g., all the references cited above). Quantities such as the energy
distribution of electrons and the ratio of electron to ion
temperatures are often taken to be constant in time and across
different regions of the accretion flow, even though neither of the
two assumptions are expected to be correct in detail.

The second reason is related to the expected and observed variability
of the emission from \sgra. The inner accretion flow is by its nature
highly turbulent and variable. As a result, even if we had a highly
predictive model for the thermodynamic properties of the plasma, it
would still be extremely unlikely that any realization of the model
would match the details of any particular snapshot of the observations
(Chan et al.\ 2015b). This would be equivalent to having a model of
cloud formation in the Earth's atmosphere and trying to fit cloud
shapes to pictures of the sky.

Because of these two reasons, inferring the properties of the
black-hole shadow as a by-product of fitting accretion flow images to
the EHT observations is not expected to be accurate. Any naive
posterior likelihood analysis will direct the model parameters towards
minimizing the variance between the observations and the incomplete
models in the high intensity regions of the image and not towards
accurately reproducing the shape and size of the black-hole shadow.

Our goal here is to study null-hypothesis tests of modified gravity
with EHT observations of the black-hole shadow around \sgra.  In order
to achieve this, we need a procedure to analyze the EHT data that
focuses on measuring directly the properties of the shadow in a manner
that is not seriously affected by our inability to predict the
brightness profile of the rest of the image. Indeed, the shape and
size of the black-hole shadow depend only on the properties of the
spacetime and can be predicted to very high accuracy without being
affected by the complexities of modeling the accretion flow.
(This is true unless the shadow is completely obscurred by
  intervening plasma or there is no background emission in the
  accretion flow on which the shadow can be cast.) As discussed in
detail below, the expected sharpness of the shadow allows us to devise
a procedure to measure accurately the properties of the shadow is
based on edge-detection schemes in image processing.

The sharpness of the shadow, i.e., the length scale over which the
image brightness declines, depends primarily on the profile of photon
emission within the flow and the velocity field. In order to explore
realistic profiles of the emission across the black-hole shadow, we
use the results of recent ray-tracing calculations on GRMHD
simulations reported in Chan et al.\ (2015a). The parameters of these
models were chosen so that model predictions agree with the
spectroscopic measurements and the 1.3~mm image size of \sgra. In Chan
et al.\ (2015a), we identified five successful models with different
black-hole spins, orientations of the observer, and physical
assumptions for the thermodynamic state of the plasma in the accretion
flow and in the jet. As a working example, we  first use here the
results of the \texttt{a9SANE} simulation that corresponds to a black
hole spinning at 90\% of the maximum spin rate, observed at 60$^\circ$
from its spin axis (see Chan et al.\ 2015a for a more detailed
discussion of this and other similar simulations and Psaltis et
al.\ 2015a for the choice of parameters).

As Figure~\ref{fig:contrast} shows, the rim of the shadow is not sharp
at the equatorial plane of the accretion flow---where there is
significant emission originating in the front side of the black
hole---and towards the Doppler boosted part of the emission. On the
other hand, the rim is significantly sharper away from the equatorial
plane and towards the Doppler deboosted part of the emission, with the
brightness of the image dropping from its maximum to its minimum value
over a displacement of $\lesssim 1$~M, producing a well defined edge.

There exist several techniques in image processing for detecting sharp
features (edges) in a model-independent fashion. Of these, two
algorithms are particularly well suited for images obtained with
interferometers: gradient methods (e.g., Marr \& Hildreth 1980; Canny
1983) and phase congruency methods (e.g., Kovesi 1999).  The gradient
method has already been applied successfully to interferometric images
to quantify the properties of the turbulent structure of the
interstellar magnetic field (Gaensler et al.\ 2011).  We will discuss
here the gradient method and explore the phase congruency method in a
future publication.

\subsection{Shadow Detection With Image Gradients}

In the gradient method of edge detection, our aim is to generate an
image not of the specific intensity of the source $I(\alpha,\beta)$,
where $\alpha$ and $\beta$ are Cartesian coordinate on the observer's
sky, but rather of the magnitude of the gradient of the specific
intensity $\vert \nabla I(\alpha, \beta)\vert$. By construction, the
image of the magnitude of the gradient will have local maxima at the
locations of the steepest gradients, which, in the case of the
expected EHT images, will coincide with the outline of the back hole
shadow.

The magnitude of the gradient of the specific intensity is given by
\begin{equation}
\vert\nabla I(\alpha,\beta)\vert=\left[
\left(\frac{\partial I}{\partial \alpha}\right)^2
+\left(\frac{\partial I}{\partial \beta}\right)^2\right]^{1/2}\;.
\end{equation}
For an image reconstructed from interferometric observations, each
partial derivative can be easily calculated from the complex
visibilities $V(u,v)$ using the van Cittert-Zernike theorem as, e.g.,
\begin{equation}
\frac{\partial I}{\partial \alpha}= \frac{2\pi i}{\lambda}
\int du \int dv\, u V(u,v) e^{2\pi i (\alpha u+\beta v)/\lambda}
\;.
\end{equation}
The above equations show that, in principle, generating a gradient
image requires adding in quadrature two images constructed after
multiplying first the complex visibilities by $iu$ and $iv$,
respectively. In practice, the construction of a gradient image
requires special care since the gradient operator in the $u-v$ plane
will amplify the measurement errors in the largest baselines. This can
be mitigated by designing an optimal high-frequency filter that will
suppress the noise without unnecessarily sacrificing
resolution. Moreover, the image reconstruction techniques that will be
used need to employ priors that do not penalize narrow features, as is
commonly done (see, e.g., Berger et al.\ 2012). In this paper, we
consider only the principles of this approach, assuming no
uncertainties and perfect coverage of the $u-v$ plane. In a follow-up
paper, we will address the issues related to applying this approach to
realistic EHT data.

Figure~\ref{fig:gradients} shows the magnitude of the gradient of the
accretion flow image shown in Figure~\ref{fig:contrast} on the
two-dimensional image plane of a distant observer as well as along
three cross sections delineated by dashed lines. As expected, the
steepest gradients and hence the brightest points in the gradient
image occur at the rim of the black-hole shadow, with a notable
exception near the equatorial plane (see also
Figure~\ref{fig:contrast}). A second, fainter and larger rim appears
at the outer edge of the image.

Having obtained a gradient image in which the rim of the black-hole
shadow appears as the most discernible feature, we can now employ a
pattern matching algorithm in order to determine the shape and size of
the shadow. we will explore here the Hough transform (Duda \& Hard
1972) and, in particular, we will follow the spirit of the closely
connected Radon transform that allows for a probabilistic description
of its results. This is crucial for our purposes because not only do
we want an algorithm to measure the properties of the black-hole
shadow but we also want to assess the statistical significance of the
results.

As discussed earlier, in General Relativity, the outline of the
black-hole shadow depends on only two parameters: the black-hole spin
and the inclination of the observer.  In principle, we can extract
both these parameters using the pattern matching technique outlined
below. In practice, however, the shape and size of the shadow has a
very weak dependence on these parameters ($\pm 4$\%, see, e.g.,
Takahashi 2004; Bozza et al.\ 2006; Johannsen 2013). Moreover, the
uncertainties in the prior knowledge of the astrophysical complexities
discussed in the previous section are comparable to or larger than the
range of shadow sizes predicted for the Kerr metric. As a result, we
will primarily be able to test whether the properties of the shadow
agree with the General Relativistic predictions and the Kerr metric
rather than measure the spin of the black hole and the orientation of
the observer. This is why we consider this approach a general
relativistic null hypothesis test.

\subsection{Shadow Pattern Matching with the Hough/Radon Transform}

The Hough transform is a pattern matching technique used in computer
vision to extract the properties of parametric curves based on an
incomplete set of measurements of their outlines (Duda \& Hard 1972). 
It is usually applied to a processed image, in which edges have been
enhanced, as is the case of the gradient image discussed above.

If we denote by $\alpha$ and $\beta$ the two Cartesian coordinates on
the image plane of a distant observer and, without lack of generality,
assume that the $\alpha$ axis is parallel to the E-W direction and
pointing towards the East, then the goal of the Hough transform is 
to identify the parameters of all curves of the form
\begin{equation}
{\cal C}(\alpha,\beta;\vec{M})=0
\end{equation}
that best match the edge structures of the image. In this expression,
we used the vector $\vec{M}$ to denote the collection of parameters
that are necessary to describe each curve. For convenience, we will
also denote the gradient image by
\begin{equation}
{\cal G}(\alpha,\beta)\equiv \left\vert \nabla I(\alpha,\beta)\right\vert\;.
\end{equation}

In the traditional Hough transform (albeit generalized to allow for
curves that are not simply line segments), one generates the Hough
accumulators (or Hough histograms) for each set of model parameters by
counting all the pixels on the gradient image that lie along the
corresponding curve for which the magnitude of the gradient is larger
than a threshold, which we will call $G_0$. In other words, the Hough
accumulators are
\begin{equation}
{\cal H}(\vec{M})=\sum_{\alpha,\beta} \left\{
\begin{array}{ll}
1, &  {\rm if}~{\cal C}(\alpha,\beta,\vec{M})=0~{\rm and}~G(\alpha,\beta)>G_0\\
0, &  {\rm otherwise}
\end{array}
\right.
\end{equation}

We can trivially generalize this definition to non pixelized images by
making use of the $\delta$- and the Heaviside functions ($H$) such
that
\begin{equation}
{\cal R}(\vec{M})=\int d\alpha \int d\beta 
\delta\left[{\cal C}(\alpha,\beta,\vec{M})\right]
H\left[G(\alpha,\beta)-G_0\right]\;.
\label{eq:HR}
\end{equation}
This last expression is the Radon transform, which is closely related
to the Hough transform and allows for a probabilistic description of
the result in terms of possible Bayesian priors (see, e.g., 
Pereira Vasconcelos 2003; Bonci et al.\ 2005). In understanding
the results of the Radon transform, special care needs to be taken in
addressing the effects of incomplete sampling and discretization of
the gradient image (see discussion in Pereira Vasconcelos 2003 and
references therein).

In order to apply the Hough/Radon transform~(\ref{eq:HR}), we need a
functional form for the shape of the black-hole shadow in terms of the
model parameters.  In General Relativity and for the Schwarzschild
metric, the shape of the black-hole shadow is a circle.  For the Kerr
metric, the shape of the shadow is known in the form of a parametric
analytic equation that, as expected, depends on only two quantities:
the black-hole spin $a$ and the inclination of the observer
$\theta_{\rm o}$ (Bardeen 1973; Chandrasekhar 1983). If we denote by
$\alpha^\prime$ and $\beta^\prime$ two orthogonal angular coordinates
on the image plane of a distant observer with $\alpha^\prime$
perpendicular to the spin axis of the black hole, the parametric form
of the shadow can be written as
  \begin{eqnarray}
\alpha^\prime(r)&=&
-m
\frac{\left[a^2 (r+1)+(r-3) r^2\right] \csc\theta_{\rm o}}{a (r-1)}
\nonumber\\
\beta_\pm^\prime(r)&=&\pm m
\frac{1}{a (r-1)}\left\{a^4 (r-1)^2 \cos ^2\theta _{\rm o}
\right.\nonumber\\
&&\quad+a (r-1) \left[a^2(r+1)+(r-3) r^2\right] 
   \cot ^2\theta _{\rm o}\nonumber\\
&&\quad\left.-r^3 \left[(r-3)^2 r-4a^2\right]\right\}^{1/2}\;.
  \label{eq:shadow}
 \end{eqnarray}
(Note some typos in Bardeen 1973, which were corrected in Chandrasekhar
  1983). In these equations, the angular coordinates have been
  normalized to the opening angle of one gravitational radius of a
  black hole of mass $M_{\rm BH}$ positioned at a distance $D$ away
  from the observer, i.e., for
\begin{equation}
m\equiv \left(\frac{GM_{\rm BH}}{Dc^2}\right)\;.
\end{equation} 
This is the quantity for which we have prior knowledge for \sgra, as
discussed in \S2.

The parameter $r$, which corresponds to the coordinate radii of the
orbits of the critical trajectories in units of $M_{\rm BH}$, takes
values in the range $r_{ph-}<r<r_{\rm ph+}$ where $r_{{\rm ph}\pm}$
are the coordinate radii of the corotating and counter rotating
equatorial photon orbits, i.e.,
\begin{equation}
r_{{\rm ph}\pm}=2\left\{1+\cos\left[\frac{2}{3}\cos^{-1}\left(\pm a\right)
\right]\right\}\;.
\end{equation}
Of course, the primed coordinate system, which is centered on the
black hole, need not be centered on or aligned with the coordinate
system on the image plane of the observer. If we denote by $\alpha_0$
and $\beta_0$ the horizontal and vertical displacements between the
origins of the two coordinate systems and by $\phi$ the angle of
rotation between the two frames (measured in degrees N of E) then
the parametric form of the rim of the black-hole shadow on the coordinate
system of the observer becomes
\begin{eqnarray}
\alpha(r)&=&\alpha_0+\alpha^\prime(r)\cos\phi+\beta_\pm^\prime(r)\sin\phi\nonumber\\
\beta_\pm(r)&=&\beta_0+\alpha^\prime(r)\sin\phi+\beta_\pm^\prime(r)\cos\phi\;.
\label{eq:param_shadow}
\end{eqnarray}

We can now use this parametric form in order to convert the surface
integral of the Hough/Radon transform~(\ref{eq:HR}) to a contour integral
over the parametric curve as
\begin{eqnarray}
&&{\cal R}(\alpha_0,\beta_0,\phi,m,a)=C\int_{r_{\rm ph-}}^{r_{\rm ph+}}dr 
\frac{ds}{dr} \nonumber\\
&&\qquad
\left\{H\left[G(\alpha,\beta_+)-G_0\right]+H\left[G(\alpha,\beta_-)-G_0\right]
\right\}\;,
\label{eq:HR_param}
\end{eqnarray}
where the expression for the derivative of the path length with respect to the
parameter $r$, i.e.,
\begin{equation}
\frac{ds}{dr}\equiv \left[ \left(\frac{d\alpha}{dr}\right)^{2}
+\left(\frac{d\beta}{dr}\right)^{2} \right]^{1/2}
\end{equation}
is given in Appendix~B. Of course, the contour integral along a curve
will be proportional to its circumference. In order not to apply
different weights to curves that have different circumferences, we
have also normalized the Radon transform by the circumference of
each curve, i.e.,
\begin{equation}
C^{-1}\equiv \int_{r_{\rm ph-}}^{r_{\rm ph+}}
\left[\left(\frac{d\alpha}{dr}\right)^{2}
  +\left(\frac{d\beta}{dr}\right)^{2} \right]^{1/2}dr\;.
\end{equation}
The Radon transform evaluated for each set of model parameters
measures the fractional length along the circumference of each curve
on which the gradient has a magnitude larger than the prescribed
threshold.

Naturally, this is not the only choice for the transform. If we have a
measure of the statistical uncertainty $\sigma_{\rm G}$ in the
measurement of each gradient, we can replace the Heaviside function
with the posterior likelihood that a particular measurement is
significant, i.e., with the error function, as
\begin{eqnarray}
&&{\cal R}(\alpha_0,\beta_0,\phi,m,a)=C\int_{r_{\rm ph-}}^{r_{\rm ph+}}dr 
\frac{ds}{dr} \nonumber\\
&&\qquad
\left\{{\rm erf}\left[\frac{G(\alpha,\beta_+)}{\sqrt{2}\sigma_{\rm G}}\right]
+{\rm erf}\left[\frac{G(\alpha,\beta_-)}{\sqrt{2}\sigma_{\rm G}}\right]
\right\}\;.
\label{eq:HR_erf}
\end{eqnarray}
Finally, given that we expect the black-hole shadow to generate the
strongest gradients in the intensity of the image, we can introduce an
additional weight to the Radon transform as
\begin{eqnarray}
&&{\cal R}(\alpha_0,\beta_0,\phi,m,a)=C\int_{r_{\rm ph-}}^{r_{\rm ph+}}dr 
\frac{ds}{dr} \nonumber\\
&&\quad
\left\{G(\alpha,\beta_+){\rm erf}\left[\frac{G(\alpha,\beta_+)}{\sqrt{2}\sigma_{\rm G}}\right]
+G(\alpha,\beta_-){\rm erf}\left[\frac{G(\alpha,\beta_-)}{\sqrt{2}\sigma_{\rm G}}\right]
\right\}\;.\nonumber\\
\label{eq:HR_w}
\end{eqnarray}
Each of these definitions have different strengths but also lead to 
Radon transforms that obey very different statistics. 

It is important to emphasize here that the interferometric
measurements of the gradients across the image will not be
uncorrelated and will not obey simple Gaussian statistics. Indeed,
given that the measurements are performed in the $u-v$ plane, all
inferred gradients in the image plane will be highly correlated. Such
correlations are not detrimental to our pattern matching algorithm,
which only requires identifying the largest, statistically significant
gradients on the image plane and does not depend strongly on the
actual value or uncertainty of each gradient. Nevertheless,
identifying the optimal definition of the Hough/Radon transform for
our purposes and exploring its statistical properties is necessary and
can be done within the context of mock simulated observations.  This
is beyond the scope of the current work and will be explored in a
future article.

\begin{figure}[t]
\psfig{file=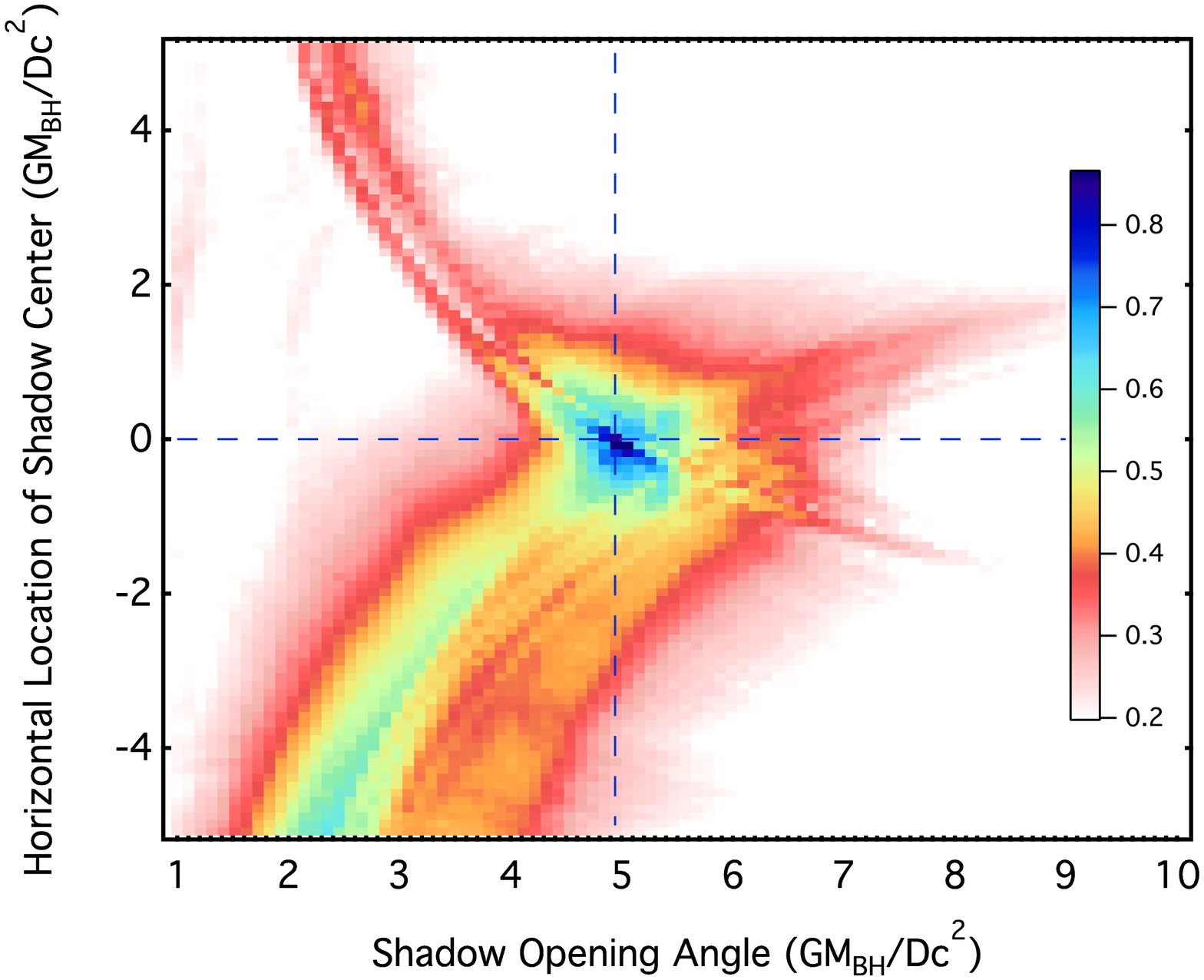,width=3.5in}
\psfig{file=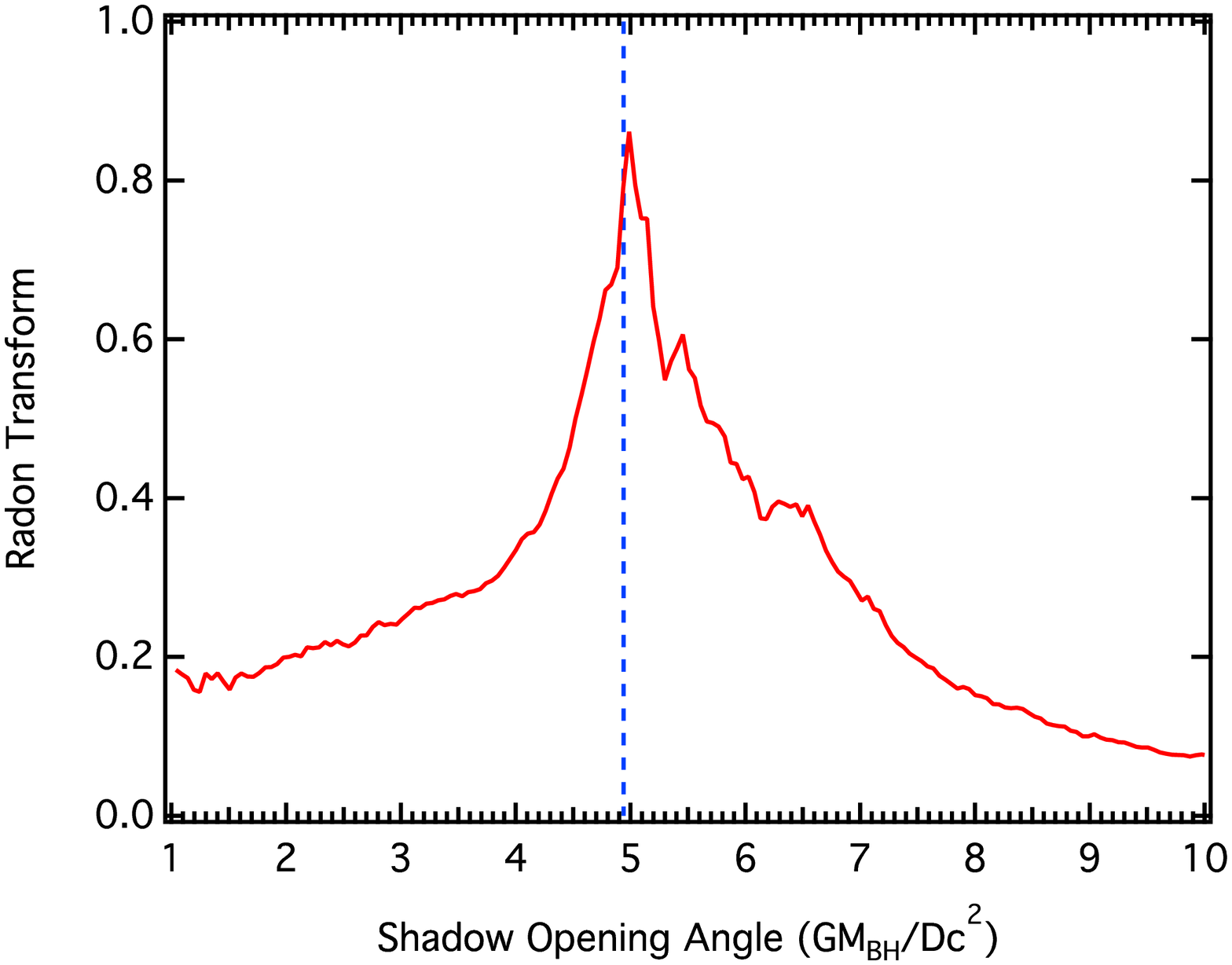,width=3.5in}
\caption{{\em (Top)\/} A two dimensional cross section of the Radon
  transform of the image shown in Figure~\ref{fig:contrast} on the
  plane of the parameter space that comprises the opening angle of the
  shadow and the horizontal location of the center of the black-hole
  shadow when the vertical location of the center of the shadow is set
  to zero ($\beta_0=0$).  {\em (Bottom)} A cross section of the Radon
  transform for black-hole shadows centered at the known location of
  the black hole shadow. The narrow prominent peak is centered at the
  expected opening angle for the simulated black hole, shown as a
  vertical dashed line, and has a fractional HWHM of 9\%. In both
  panels, all quantities are normalized to the ratio $GM_{\rm
    BH}/Dc^2$.}
\label{fig:radon}
\end{figure}

The parametric expressions~(\ref{eq:param_shadow}) for the black-hole
shadow are valid only for non-zero values of the black-hole spin. In
the special case of a Schwarzschild black hole, the rim of the shadow
is a circle with an apparent radius of $3\sqrt{3}m$, independent of the
orientation of the observer. In order to account for this special case, we 
also need a different expression for the Radon transform. The 
parametric equation for rim of the shadow on the image plane of the observer
become
\begin{eqnarray}
\alpha(\Omega)&=&\alpha_0+3\sqrt{3}m \cos\Omega\\
\beta_\pm(\Omega)&=&\beta_0\pm 3\sqrt{3}m \sin\Omega\;,
\end{eqnarray}
with $0\le\Omega<\pi$.  The Radon transform, in this case, then simply
becomes a function of the three model parameters
\begin{eqnarray}
&&{\cal R}(\alpha_0,\beta_0,m)=\frac{1}{2\pi}\int_0^\pi d\Omega
  \nonumber\\ &&\qquad
  \left\{H\left[G(\alpha,\beta_+)-G_0\right]+H\left[G(\alpha,\beta_-)-G_0\right]
  \right\},
\label{eq:HR_Sch}
\end{eqnarray}
For the examples in this paper, we will use this last version of the
Radon transform, i.e., assuming that the size of the black-hole shadow
is always circular and equal to that of a Schwarzschild black
hole. This is justified given the weak dependence of the black-hole
size on black-hole spin and observer inclination (see, e.g., Chan et
al.\ 2013) and the approximations regarding the observations that we
will employ below. It is worth emphasizing, however, that there is no
inherent difficulty in employing the complete form of the Radon
transform and marginalizing over the spin of the black hole and the
inclination of the observer.

The Hough/Radon transform~(\ref{eq:HR_Sch}) is a three-dimensional
function of the model parameters $\alpha_0$, $\beta_0$, and $m$. Two
of them are nuisance parameters (the coordinates of the center of the
image), whereas the third parameter $m$ is the one of interest. The
top panel of Figure~\ref{fig:radon} shows a cross section of the
three-dimensional Hough/Radon transform for the gradient image of
Figure~\ref{fig:gradients}. The vertical dashed line corresponds to
the known opening angle (in units of $GM/Dc^2$) of the black-hole
shadow. 

The bottom panel of Figure~\ref{fig:radon} shows the Hough/Radon
transform, along a cross section of the parameter space that
corresponds to $\alpha_0=\beta_0=0$, i.e., for curves centered at the
known center of the black-hole shadow. The peak occurs at the expected
size of the black-hole shadow and its HWHM is approximately equal to
9\%. This is the limiting accuracy with which the sharp gradient at
the rim of the shadow can be localized, based on the most likely
properties of the plasma emission in the vicinity of \sgra, as
inferred from our GRMHD simulations.

\begin{figure}[t]
\psfig{file=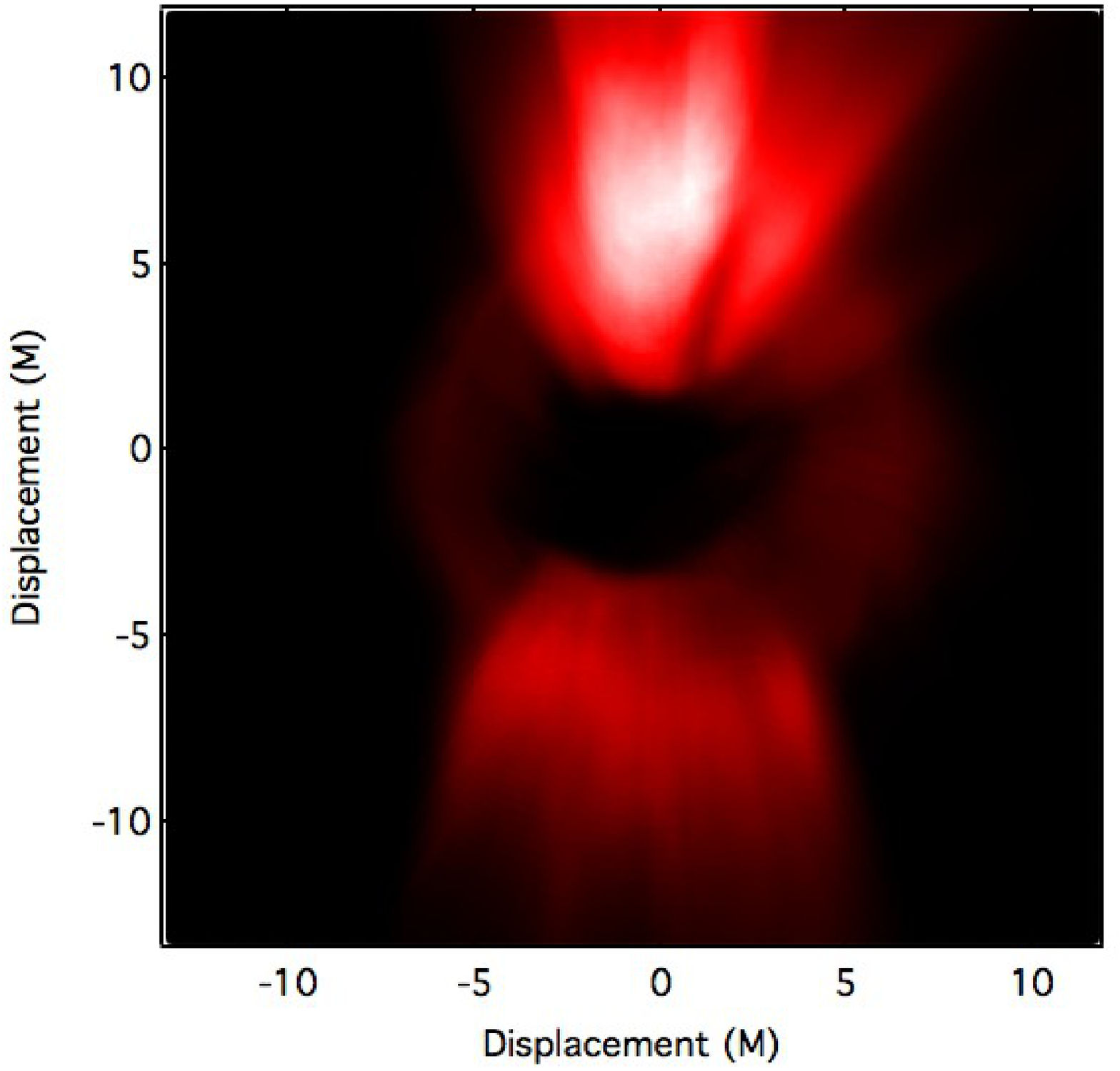,width=3.5in,height=3.0in}
\psfig{file=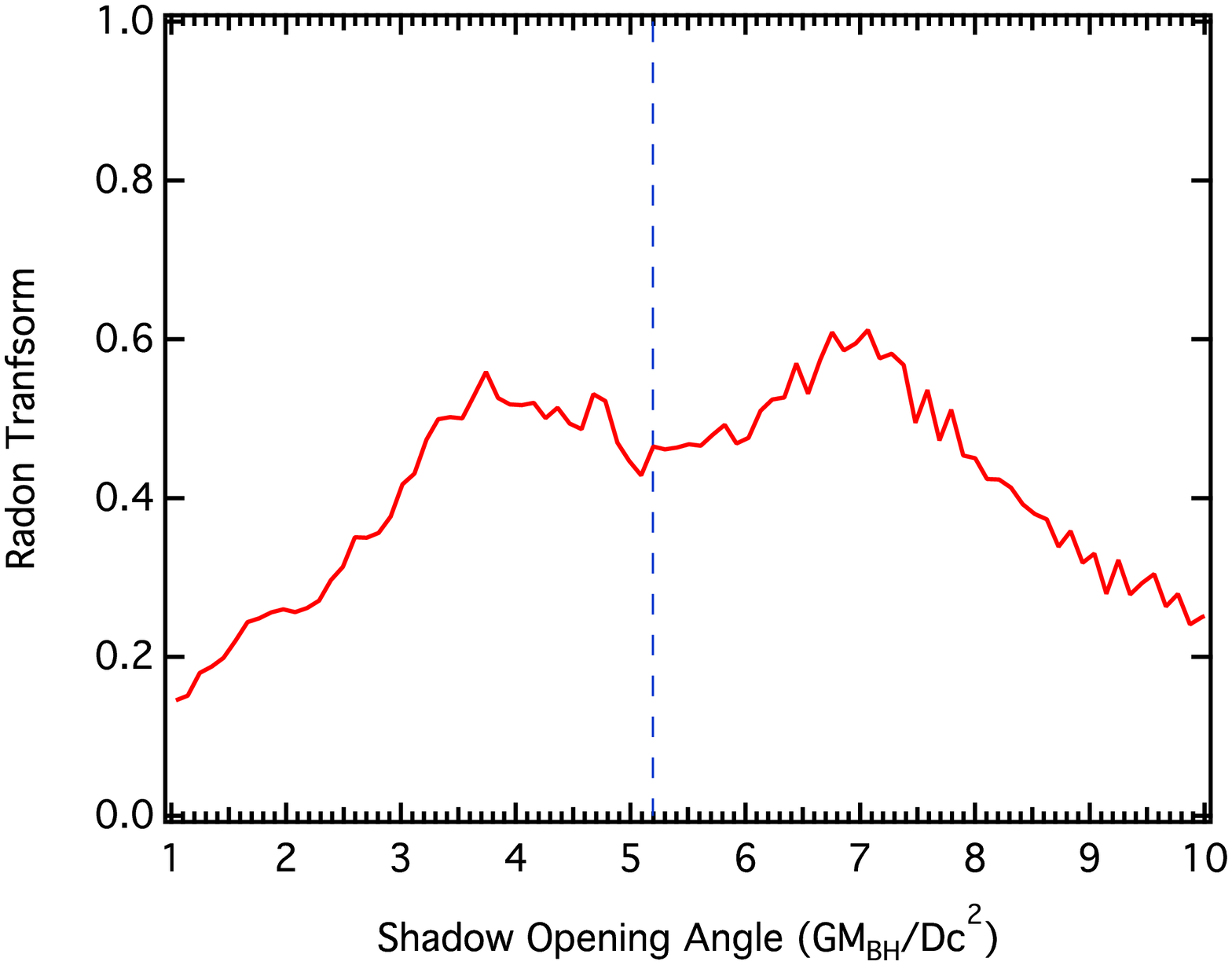,width=3.5in}
\caption{{\em (Top)\/} The simulated image for a MAD accretion flow
  around \sgra~(\texttt{a0MAD} simulation in Chan et al.\ 2015a). This
  image is dominated by emission from the footpoints of the jets and
  shows only a very weak black-hole shadow. {\em (Bottom)\/} A cross
  section of the Radon transform for this image, for a pattern
  centered at the known location of the black hole shadow. When the
  emission pattern, or other astrophysical complications, do now allow
  for a signficant fraction of the black-hole shadow to become
  prominant, the Radon transform is inconclusive and does not lead to
  a biased measurement of the shadow properties.}
\label{fig:mad}
\end{figure}

The particular simulation that we have used in the previous example
shows a prominent shadow that shapes the brightness distribution of
the images of the accretion flow. Is is possible, of course, that the
images are dominated instead by bright spots with very weak evidence
for the sharp gradients at the rim of the black-hole shadow.  This is
true, e.g., for images dominated by emission at the footpoints of
jets, as in the simulations of Moscibrodzka et al.\ (2014) and the
\texttt{MAD} simulations of Chan et al.\ (2015a), or at the standing
shocks of tilted accretion disks, as in the simulations of Dexter \&
Fragile (2013).  Alternatively, the intervening plasma might be blocking
partially or fully the shadow and/or other structures, such as
gravitationally lensed images of sheared flux tubes (see, e.g.,
Figure~9 of Chan et al.\ 2015b), might be mistaken for the shadow
outline. In the latter case, such structures would be very short lived
and repeated applications of the technique discussed here to images
observed at different epochs should be able to filter against them. In
the former case, however, it is important to show that the Radon
transform of the gradient image will not lead to a biased measurement
of the shadow properties.

The Radon transform becomes highly peaked only if sharp image
gradients appear along a significant fraction of the pattern that is
searched for. If the image shows shallower gradients along patterns
that do not follow the expected outline of a Kerr shadow, then the
Radon transform will not lead to a statistically significant
measurement. To demonstrate this, Figure~\ref{fig:mad} shows the
time-averaged image of the \texttt{a0MAD} simulation of Chan et
al.\ (2015a), which is dominated by emission at the footpoints of the
jet. The Radon transform of the gradient image in this case is very
broad and inconclusive on whether the black-hole shadow has been
detected or not. Determining the properties of the shadow from an image
such as the one shown in this figure would be very challenging.
Nevertheless, the advantage of the Radon transform is that it does not
pick an erroneous value for the opening angle of the shadow but simply
remains inconclusive. Astrophysical complexities might prohibit but
will not bias a null hypothesis test of general relativity, as the one
we discuss here.

\section{Conclusions}

The anticipated EHT observations of the black-hole shadow in
\sgra\ will provide some of the strongest astrophysical evidence for
the existence of black holes in the Universe and a direct observation
of the effects of strong-field gravitational lensing. In this paper,
we explored the prospect of performing a {\em qualitative\/} test of
General Relativity using these observations in a manner that does not
depend on the complex modeling of the accretion flow. In particular,
we investigated the prospect of verifying the General Relativistic
prediction for the half opening angle of the black-hole shadow in the
Kerr metric, which is expected to be within 4\% of $5 GM/Dc^2$, for all
black-hole spins and orientations of the distant observer.

Performing this null-hypothesis test requires four distinct steps.
First, the theoretical prediction for the linear size of the
black-hole shadow needs to be converted to an angular size on the
observer's sky. This requires prior knowledge of the mass-to-distance
ratio for \sgra, which is currently known to an adequate accuracy
($\sim 4$\%) from monitoring of stellar orbits. Future observations
with next generation adaptive optics in 30-m class telescopes will
reduce this uncertainty to less than one part in a thousand (Weinberg
et al.\ 2005), effectively removing it from the error budget of the
test.

Second, the blurring effects of scattering in the turbulent
interstellar medium need to be removed in order for the sharp
black-hole shadow to be easily distinguishable from the surrounding
emission. Current observations suggest that the measurements of the
scattering kernel need to be performed at large wavelengths (10-20~cm)
and then extrapolated down to the 1.3~mm wavelength of EHT
observations. Even though current measurements of the properties of
the scattering ellipse at $10-20$~cm are accurate to $3-20$\%,
extrapolating them requires a theoretical model for the wavelength
dependence of scattering induced blurring in anisotropic turbulence.
This is the most severe current theoretical limitation in the null
hypothesis test of General Relativistic that we propose here.

Third, the rim of the black hole shadow needs to be identified in the
interferometric images, in a manner that does not depend on our prior
knowledge and modeling of the accretion flow. We explored here an edge
detection algorithm that identifies the black-hole shadow with the
location of the sharpest gradients in the brightness image. Similar
algorithms based on phase congruency are also likely to be applicable.
Applying such algorithms directly to EHT data requires a detailed
study of the statistics of the measured gradients, which will be
highly correlated and non Gaussian.

Fourth, a pattern matching algorithm needs to be applied to the
gradient images that will measure the properties of the black-hole
shadow and in particular its opening angle. To this end, we
investigated a variant of the Hough/Radon transform and applied it to
mock images calculated from GRMHD simulations that have been tuned to
match all current observations of \sgra. We found that, if
\sgra\ behaves according to the most detailed currently available
numerical models, then the Hough/Radon transform can lead to measurements
of the opening angle of the black-hole shadow with an accuracy that is
$\lesssim 10$\%.

Our main goal in this work has been to explore the principles of
performing this quantitative null hypothesis test of General
Relativity with EHT observations of \sgra. Our results now warrant a
more detailed investigation of the practical application of our
approach, considering the sparse sampling of the interferometric $u-v$
plane, the correlated uncertainties in the measurements, and the
characteristic of the image reconstruction algorithms that will be
used. We will explore these in a forthcoming article.

\acknowledgements

We thank Andrea Ghez, Sylvana Yelda, and Stefan Gillessen for
providing us with their measurements of the mass and distance to
\sgra\ based on the monitoring of stellar orbits. DP, FO, and CC were
partially supported by NASA TCAN award NNX14AB48G. DP and CC also
acknowledge support from NSF grant AST~1312034. FO ackowledges support
from the Miller Institute for Basic Research at UC Berkeley.  DM
acknowledges support from NSF grant AST-1207752. All ray tracing
calculations were performed with the \texttt{El Gato} GPU cluster at
the University of Arizona that is funded by NSF award 1228509.

\appendix

\section{A.\ A Bayesian Analysis of the Properties of the Scattering Screen}

In our investigation of the properties of the scattering screen
towards \sgra, we faced the problem of combining measurements that are
not necessarily in statistical agreement with each other. Such
measurements suggest the presence of systematic unertainties that have
not been accounted for in their error budget. If these systematic
uncertainties are unaccounted for, they will lead to a significant
underestimate of the final uncertainties of the properties of the
scattering screen. In this appendix, we present a Bayesian analysis
based on our earlier work (G\"uver et al.\ 2012a, 2012b) that allows
us to quantify the degree of systematic uncertainties in an ensemble
of measurements.

We will assume that each measurement of a quantity $f$ in the ensemle
is drawn from a Gaussian distribution with a centroid at
$f_0$ and a dispersion $\sigma$, i.e.,
\begin{equation}
P(f;f_0,\sigma)=C\exp\left[-\frac{(f-f_0)^2}{2\sigma^2}\right]\;,
\end{equation}
where $C$ is a proper normalization constant. This distribution is our
model for the underlying systematic uncertainties in the measurements.
Our goal is to calculate the posterior likelihood for the parameters
$f_0$ and $\sigma$, $P(f_0,\sigma|$data$)$, given an ensemble of $N$
measurements. Each of the measurements $f_i$, is described by a
posterior likelihood, which we will assume here to be also a Gaussian
with a centroid $f_{0,i}$ and a disperssion $\sigma_{i}$ i.e.,
\begin{equation}
P_i(f|f_{0,i},\sigma_i)=C_i\exp\left[-\frac{(f-f_{0,i})^2}{2\sigma_i^2}\right]\;,
\end{equation}
where $C_i$ is a set of normalization constants.

Using Bayes' theorem, we can write 
\begin{equation}
P(f_0,\sigma|\mbox{data})=C_0 P(\mbox{data}|f_0,\sigma) P(f_0)P(\sigma)\;.
\label{eq:bayes}
\end{equation}
Here $C_0$ is a normalization constant, and $P(f_0)$ and $P(\sigma)$
are the priors over the centroid and dispersion of the underlying
Gaussian distribution, which we take to be flat in a range larger than
the anticipated values. Figure~(\ref{fig:normalized}) shows that
  the $1-$ and $2-\sigma$ contours for the centroids of the three
  parameters of interest are quite narrow and, therefore, different
  broad prior distributions would have a very small effect on the
  final results. The range of statistically allowed dispersions, on
  the other hand, are larger and, therefore, are more susceptible to
  the properties of the prior distribution. When additional data are
  incorporated in this analysis to improve the statistics of the
  measured parameters, the effect of different priors will need to be
  investigated more thoroughly.

The quantity $P(\mbox{data}|f_0,\sigma)$ measures the likelihood that
we will make a particular set of measurements given the values of the
parameters of the underlying Gaussian distribution. Assuming for
simplicity that these measurements are uncorrelated, we write
\begin{equation}
P(\mbox{data}|f_0,\sigma)=\prod_i \int df P_i(f|f_{0,i},\sigma_i)
P(f;f_0,\sigma)\;.
\label{eq:data}
\end{equation}

Inserting equation~(\ref{eq:data}) into equation~(\ref{eq:bayes}) we
obtain the sought after posterior likelihood
\begin{equation}
P(f_0,\sigma|\mbox{data})=C_0 P(f_0)P(\sigma)
\prod_i \int df P_i(f|f_{0,i},\sigma_i)
P(f;f_0,\sigma)\;,
\end{equation}
where $C_0$ is an overall normalization constant. 

Finally, in order to compute the posterior likelihood for the quantity $f$
given the two dimensional posterior likelihood of the centroid value $f_0$
and dispersion $\sigma$ of the underlying Gaussian distribution, we write
\begin{equation}
P(f)=\int \int P(f;f_0,\sigma) P(f_0,\sigma|\mbox{data}) df_0 d\sigma\;.
\end{equation}

\section{B.\ The contour integral in the Radon transform}

In the calculation of the Radon transform using a parametric curve for
the black-hole shadow, we need the derivative of the path length along
the rim of the black-hole shadow and the parameter $r$. For a Kerr
black hole, this is given by
\begin{eqnarray}
\frac{ds}{dr}&=&
\frac{r \left(r^2-3 r+3\right)-a^2}
{a (r-1)^2}\nonumber\\
&&
\left\{\frac{\left[a(r-1) \cot ^2\theta _{\rm o}-2 (r-3) r^2\right]^2}
            {a^4 (r-1)^2 \cos^2\theta_{\rm o}
              +a (r-1) \left[a^2 (r+1)+(r-3) r^2\right] \cot^2\theta_{\rm o}
              -r^3 \left[(r-3)^2 r-4 a^2\right]}+4 \csc^2\theta_{\rm o}\right\}^{1/2}\;.
\end{eqnarray}


\begin{thebibliography}{}

\bibitem[Amarilla \& Eiroa(2013)]{2013PhRvD..87d4057A} Amarilla, L.,
  \& Eiroa, E.~F.\ 2013, \prd, 87, 044057

\bibitem[Akiyama et al.(2013)]{2013PASJ...65...91A} Akiyama, K., Takahashi, 
R., Honma, M., Oyama, T., \& Kobayashi, H.\ 2013, \pasj, 65, 91 

\bibitem[Alberdi et 
al.(1993)]{1993A&A...277L...1A} Alberdi, A., Lara, L., Marcaide, J.~M., et al.\ 1993, \aap, 277, L1 

\bibitem[An et al.(2005)]{2005ApJ...634L..49A} An, T., Goss, W.~M., Zhao, 
J.-H., et al.\ 2005, \apjl, 634, L49 

\bibitem[Bambi \& Yoshida(2010)]{2010CQGra..27t5006B} Bambi, C., \&
  Yoshida, N.\ 2010, Classical and Quantum Gravity, 27, 205006

\bibitem[Bambi \& Freese(2009)]{2009PhRvD..79d3002B} Bambi, C., \&
  Freese, K.\ 2009, \prd, 79, 043002

\bibitem[Bardeen et al.(1972)]{1972ApJ...178..347B} Bardeen, J.~M., Press, 
W.~H., \& Teukolsky, S.~A.\ 1972, \apj, 178, 347 

\bibitem[Bardeen(1973)]{1973blho.conf..215B} Bardeen, J.~M.\ 1973, Black 
Holes (Les Astres Occlus), 215 

\bibitem[Berger et al.(2012)]{2012A&ARv..20...53B} Berger, J.-P.,
  Malbet, F., Baron, F., et al.\ 2012, \aapr, 20, 53

\bibitem[Boehle et al.(2015)]{bb} Boehle, et al.\ 2015, submitted
  
\bibitem[Bonci et al.(2005)]{2005IEEE} Bonci, A., Leo, T., \& Longhi, S.\
2005, IEEE Trans.\ Sys.\ Man.\ Cyb., A 35, 945

\bibitem[Bower et al.(2004)]{2004Sci...304..704B} Bower, G.~C., Falcke, H., 
Herrnstein, R.~M., et al.\ 2004, Science, 304, 704 

\bibitem[Bower et al.(2006)]{2006ApJ...648L.127B} Bower, G.~C., Goss, 
W.~M., Falcke, H., Backer, D.~C., \& Lithwick, Y.\ 2006, \apjl, 648, L127 

\bibitem[Bower et al.(2014)]{2014ApJ...790....1B} Bower, G.~C., Markoff, 
S., Brunthaler, A., et al.\ 2014, \apj, 790, 1 

\bibitem[Bozza et al.(2006)]{2006PhRvD..74f3001B} Bozza, V., de Luca, F., 
\& Scarpetta, G.\ 2006, \prd, 74, 063001 

\bibitem[Broderick et al.(2009)]{2009ApJ...697...45B} Broderick, A.~E., 
Fish, V.~L., Doeleman, S.~S., \& Loeb, A.\ 2009, \apj, 697, 45 

\bibitem[Broderick et al.(2011)]{2011ApJ...735..110B} Broderick, A.~E., 
Fish, V.~L., Doeleman, S.~S., \& Loeb, A.\ 2011, \apj, 735, 110 

\bibitem[Canny(1983)]{1983mit..reptQ....C} Canny, J.~F.\ 1983,
  M.Sc.\ Thesis, Massachusetts Inst.~of Tech.~Report, 720

\bibitem[Chan et al.(2013)]{2013ApJ...777...13C} Chan, C.-k., Psaltis,
  D., \"Ozel, F.\ 2013, \apj, 777, 13

\bibitem[Chan et al.(2014)]{2014arXiv1410.3492C} Chan, C.-K., Psaltis,
  D., Ozel, F., Narayan, R., \& Sadowski, A.\ 2015a, \apj, 799, 1

  \bibitem[Chan et al.(2015)]{2015arXiv150501500C} Chan, C.-k.,
    Psaltis, D., Ozel, F., et al.\ 2015b, ApJ, in press,
    arXiv:1505.01500

\bibitem[Chandran \& Backer(2002)]{2002ApJ...576..176C} Chandran,
  B.~D.~G., \& Backer, D.~C.\ 2002, \apj, 576, 176

\bibitem[Chandrasekhar(1983)]{1983mtbh.book.....C} Chandrasekhar,
  S.\ 1983, The Mathematical Theory of Black Holes, Oxford/New York,
  Clarendon Press/Oxford University Press (International Series of
  Monographs on Physics.~Volume 69)

\bibitem[de Vries(2000)]{2000CQGra..17..123D} de Vries, A.\ 2000,
  Classical and Quantum Gravity, 17, 123

\bibitem[Dexter et al.(2009)]{2009ApJ...703L.142D} Dexter, J., Agol, E., 
\& Fragile, P.~C.\ 2009, \apjl, 703, L142 

\bibitem[Dexter et al.(2010)]{2010ApJ...717.1092D} Dexter, J., Agol, E., 
Fragile, P.~C., \& McKinney, J.~C.\ 2010, \apj, 717, 1092 

\bibitem[Dexter \& Fragile(2013)]{2013MNRAS.432.2252D} Dexter, J., \&
  Fragile, P.~C.\ 2013, \mnras, 432, 2252

\bibitem[Doeleman et al.(2008)]{2008Natur.455...78D} Doeleman, S.~S.,
  Weintroub, J., Rogers, A.~E.~E., et al.\ 2008, \nat, 455, 78

\bibitem[Duda \& Hard(1972)]{1972} Duda, R.\,O.\ \& Hard,
  P.\,E.\ 1972, Communications of the ACM, 15, 11

\bibitem[Eisenhauer et al. (2011)]{2011Msngr.143...16E} Eisenhauer,
  F., Perrin, G., Brandner, W., et al.\ 2011, The Messenger, 143, 16

\bibitem[Falcke 
\& Markoff(2013)]{2013CQGra..30x4003F} Falcke, H., \& Markoff, S.~B.\ 2013, Classical and Quantum Gravity, 30, 244003 

\bibitem[Falcke et al.(2000)]{2000ApJ...528L..13F} Falcke, H., Melia, F., 
\& Agol, E.\ 2000, \apjl, 528, L13 

\bibitem[Fish et al.(2009)]{2009ApJ...692L..14F} Fish, V.~L.,
  Broderick, A.~E., Doeleman, S.~S., \& Loeb, A.\ 2009, \apjl, 692,
  L14

\bibitem[Fish et al.(2014)]{2014ApJ...795..134F} Fish, V.~L., Johnson, 
M.~D., Lu, R.-S., et al.\ 2014, \apj, 795, 134 

\bibitem[Gaensler et al.(2011)]{2011Natur.478..214G} Gaensler, B.~M., 
Haverkorn, M., Burkhart, B., et al.\ 2011, \nat, 478, 214 

\bibitem[Ghez et al.(2008)]{2008ApJ...689.1044G} Ghez, A.~M., Salim, S., 
Weinberg, N.~N., et al.\ 2008, \apj, 689, 1044 

\bibitem[Gillessen et al.(2009)]{2009ApJ...707L.114G} Gillessen, S., 
Eisenhauer, F., Fritz, T.~K., et al.\ 2009a, \apjl, 707, L114 

\bibitem[Gillessen et al.(2009)]{2009ApJ...692.1075G} Gillessen, S., 
Eisenhauer, F., Trippe, S., et al.\ 2009b, \apj, 692, 1075 

\bibitem[Goodman \& Narayan(1989)]{1989MNRAS.238..995G} Goodman, J.,
  \& Narayan, R.\ 1989, \mnras, 238, 995

\bibitem[G{\"u}ver et al.(2012)]{2012ApJ...747...76G} G{\"u}ver, T., 
Psaltis, D., {\"O}zel, F.\ 2012a, \apj, 747, 76 

\bibitem[G{\"u}ver et al.(2012)]{2012ApJ...747...77G} G{\"u}ver, T., 
{\"O}zel, F., \& Psaltis, D.\ 2012b, \apj, 747, 77 

\bibitem[Gwinn et al.(2014)]{2014ApJ...794L..14G} Gwinn, C.~R., Kovalev, 
Y.~Y., Johnson, M.~D., \& Soglasnov, V.~A.\ 2014, \apjl, 794, LL14 

\bibitem[Jaroszynski \& Kurpiewski(1997)]{1997A&A...326..419J}
  Jaroszynski, M., \& Kurpiewski, A.\ 1997, \aap, 326, 419

\bibitem[Jauncey et al.(1989)]{1989AJ.....98...44J} Jauncey, D.~L., 
Tzioumis, A.~K., Preston, R.~A., et al.\ 1989, \aj, 98, 44 

\bibitem[Johannsen(2013)]{2013PhRvD..87l4017J} Johannsen, T.\ 2013, \prd, 
87, 124017 

\bibitem[Johannsen \& Psaltis(2010)]{2010ApJ...718..446J} Johannsen,
  T., \& Psaltis, D.\ 2010, \apj, 718, 446

\bibitem[Johnson \& Gwinn(2015)]{2015ApJ...805..180J} Johnson, M.~D.,
  \& Gwinn, C.~R.\ 2015, \apj, 805, 180
  
\bibitem[Kovesi(1999)]{1999Kovesi} Kovesi, P.\,D.\ 1999, Journal of
  Computer Vision Research 1

\bibitem[Liu et al.(2012)]{2012ApJ...747....1L} Liu, K., Wex, N.,
  Kramer, M., Cordes, J.~M., \& Lazio, T.~J.~W.\ 2012, \apj, 747, 1

\bibitem[Lo et al.(1993)]{1993Natur.362...38L} Lo, K.~Y., Backer, D.~C., 
Kellermann, K.~I., et al.\ 1993, \nat, 362, 38 

\bibitem[Lo et al.(1985)]{1985Natur.315..124L} Lo, K.~Y., Backer, D.~C., 
Ekers, R.~D., et al.\ 1985, \nat, 315, 124 

\bibitem[Lu et al.(2011)]{2011A&A...525A..76L} Lu, R.-S., Krichbaum,
  T.~P., Eckart, A., et al.\ 2011, \aap, 525, A76

\bibitem[Luminet(1979)]{1979A&A....75..228L} Luminet, J.-P.\ 1979,
  \aap, 75, 228

\bibitem[Marcaide et al.(1999)]{1999A&A...343..801M} Marcaide, J.~M.,
  Alberdi, A., Lara, L., P{\'e}rez-Torres, M.~A., \& Diamond,
  P.~J.\ 1999, \aap, 343, 801

\bibitem[Marr \& Hildreth(1980)]{1980RSPSB.207..187M} Marr, D., \&
  Hildreth, E.\ 1980, Royal Society of London Proceedings Series B,
  207, 187

\bibitem[Mo{\'s}cibrodzka et al.(2009)]{2009ApJ...706..497M}
  Mo{\'s}cibrodzka, M., Gammie, C.~F., Dolence, J.~C., Shiokawa, H.,
  \& Leung, P.~K.\ 2009, \apj, 706, 497

\bibitem[Mo{\'s}cibrodzka \& Falcke(2013)]{2013A&A...559L...3M}
  Mo{\'s}cibrodzka, M., \& Falcke, H.\ 2013, \aap, 559, LL3

\bibitem[Mo{\'s}cibrodzka et al.(2014)]{2014A&A...570A...7M}
  Mo{\'s}cibrodzka, M., Falcke, H., Shiokawa, H., \& Gammie,
  C.~F.\ 2014, \aap, 570, AA7

\bibitem[Narayan \& Goodman(1989)]{1989MNRAS.238..963N} Narayan, R.,
  \& Goodman, J.\ 1989, \mnras, 238, 963

\bibitem[{\"O}zel et al.(2000)]{2000ApJ...541..234O} {\"O}zel, F., Psaltis, 
D., \& Narayan, R.\ 2000, \apj, 541, 234

\bibitem[Pereira(2003)]{2003PhDT} Pereira Vasconcelos, M.\,A.,S.\ 2003, Ph.D.~Thesis, Harvard University

\bibitem[Pfahl \& Loeb(2004)]{2004ApJ...615..253P} Pfahl, E., \& Loeb,
  A.\ 2004, \apj, 615, 253

\bibitem[Psaltis et al.(2015)]{2015ApJ...798...15P} Psaltis, D., Narayan, 
R., Fish, V.~L., et al.\ 2015a, \apj, 798, 15 
  
\bibitem[Psaltis (2015)]{P15} Psaltis, D., Wex, N., \& Kramer, M.\ 2015b,
  \apj, in press
  
\bibitem[Roy \& Pramesh Rao(2003)]{2003ANS...324..391R} Roy, S., \&
  Pramesh Rao, A.\ 2003, Astronomische Nachrichten Supplement, 324,
  391

\bibitem[Shen et al.(2005)]{2005Natur.438...62S} Shen, Z.-Q., Lo, K.~Y., 
Liang, M.-C., Ho, P.~T.~P., \& Zhao, J.-H.\ 2005, \nat, 438, 62 

\bibitem[Takahashi(2004)]{2004ApJ...611..996T} Takahashi, R.\ 2004, \apj, 
611, 996 

\bibitem[Weinberg et al.(2005)]{2005ApJ...622..878W} Weinberg, N.~N.,
  Milosavljevi{\'c}, M., \& Ghez, A.~M.\ 2005, \apj, 622, 878

\bibitem[Yusef-Zadeh et al.(1994)]{1994ApJ...434L..63Y} Yusef-Zadeh, F., 
Cotton, W., Wardle, M., Melia, F., \& Roberts, D.~A.\ 1994, \apjl, 434, L63 

\end{thebibliography}
\end{document}